%% file: DistributedNetworkAnalysis-arxiv.tex
\documentclass[twocolumn]{article}
\pdfoutput=1

\usepackage[margin=1.8cm]{geometry}
\usepackage[T1]{fontenc}
\usepackage[utf8]{inputenc}
\usepackage{hyperref}
\usepackage{float}
\usepackage{url}
\usepackage{graphicx}
\usepackage{setspace}
\usepackage{tikz}
\usepackage{pgfplots}
\usepackage{standalone}
\usepackage{amssymb}
\usepackage{epstopdf}
\usepackage{rotating}
\usepackage{subfig}
\usepackage{dblfloatfix} 
\usepackage{float} 
\usepackage{fixltx2e} 
\usepackage{soul} 
\usepackage{multirow}
\usepackage{todonotes}
\usepackage{wrapfig}
\usepackage{rotating}
\usepackage{xcolor}
\usepackage{fancyhdr}
\usepackage{array}
\usepackage{numprint}
\usepackage{xspace}
\usepackage[switch]{lineno}
\usepackage{xr}	

\newcommand{\ie}{i.\,e.\xspace}
\newcommand{\eg}{e.\,g.\xspace}

\newcommand{\algpr}{\texttt{PAGERANK-FIXED}\xspace}

\newcommand{\algcc}{\texttt{CON-COMP}\xspace}
\newcommand{\algccm}{\texttt{CON-COMP-MSG}\xspace}

\input{./codes/pseudocodes_setup}
\input{./img/plots/plots_setup}

\definecolor{bg}{rgb}{0.985,0.985,0.985}

\hypersetup{breaklinks=true,
            pdfauthor={Jannis Koch, Christian L. Staudt, Maximilian Vogel, Henning Meyerhenke},
            pdftitle={An Empirical Comparison of Big Graph Frameworks in the Context of Network Analysis}
            colorlinks=true,
            citecolor=PineGreen,
            urlcolor=MidnightBlue,
            linkcolor=PineGreen,
            pdfborder={0 0 0}}
\definecolor{bg}{rgb}{0.985,0.985,0.985}

\title{An Empirical Comparison of Big Graph Frameworks in the Context of Network Analysis}
\author{Jannis Koch \and Christian L. Staudt \and Maximilian Vogel \and Henning Meyerhenke}
\date{}                                           

\begin{document}
\maketitle

\begin{abstract} 
Complex networks are relational data sets commonly represented as graphs.
The analysis of their intricate structure is relevant to many areas of science and commerce, and data sets may reach sizes that require distributed storage and processing.

We describe and compare programming models for distributed computing with a focus on graph algorithms for large-scale complex network analysis.
Four frameworks -- GraphLab, Apache Giraph, Giraph++ and Apache Flink -- are used to implement algorithms for the representative problems Connected Components, Community Detection, PageRank  and Clustering Coefficients.
The implementations are executed on a computer cluster to evaluate the frameworks’ suitability in practice
 and to compare their performance to that of the single-machine, shared-memory parallel network analysis package NetworKit.
 Out of the distributed frameworks, GraphLab and Apache Giraph generally show the best performance. 

In our experiments a cluster of eight computers running Apache Giraph enables the analysis of a network with about 2 billion edges, which is too large for a single machine of the same type.
However, for networks that fit into memory of one machine, the performance of the shared-memory parallel implementation is far better than the distributed ones. The study provides experimental evidence for selecting the appropriate framework depending on the task and data volume.

\small\baselineskip=9pt ~\\[0.5ex]\noindent \textbf{Keywords: big graph frameworks, distributed computing, graph algorithms, complex networks, network analysis} 
\end{abstract}

\section{Introduction}

 
Complex networks are data sets which map items and the links among them, forming complex relational patterns.
They are typically represented as graphs, i.e. a set of $n$ vertices $V$ and $m$ edges $E$.
This abstraction is as general as it is powerful, and has been employed in numerous problem domains~\cite{costa2011analyzing}.
The task of network analysis is to characterize the structure of the network as a whole or identify important elements in it, typically by applying a collection of graph algorithms tailored to this purpose (see Section~\ref{sec:algorithms}).

Many applications -- think for example of the networks collected by online social networking sites -- produce enormous amounts of network data that exceed the memory of any single computer, calling for distributed solutions.
A distributed system consists of a set of autonomous processors, each with their own memory, running software that performs computations and exchanges data as messages via a network.
Distributed systems are highly scalable in the sense that additional computers can be added easily. Of course, additional computers may
mean additional overhead as well.

\begin{figure}[htb]
\begin{center}
\includegraphics[width=\columnwidth]{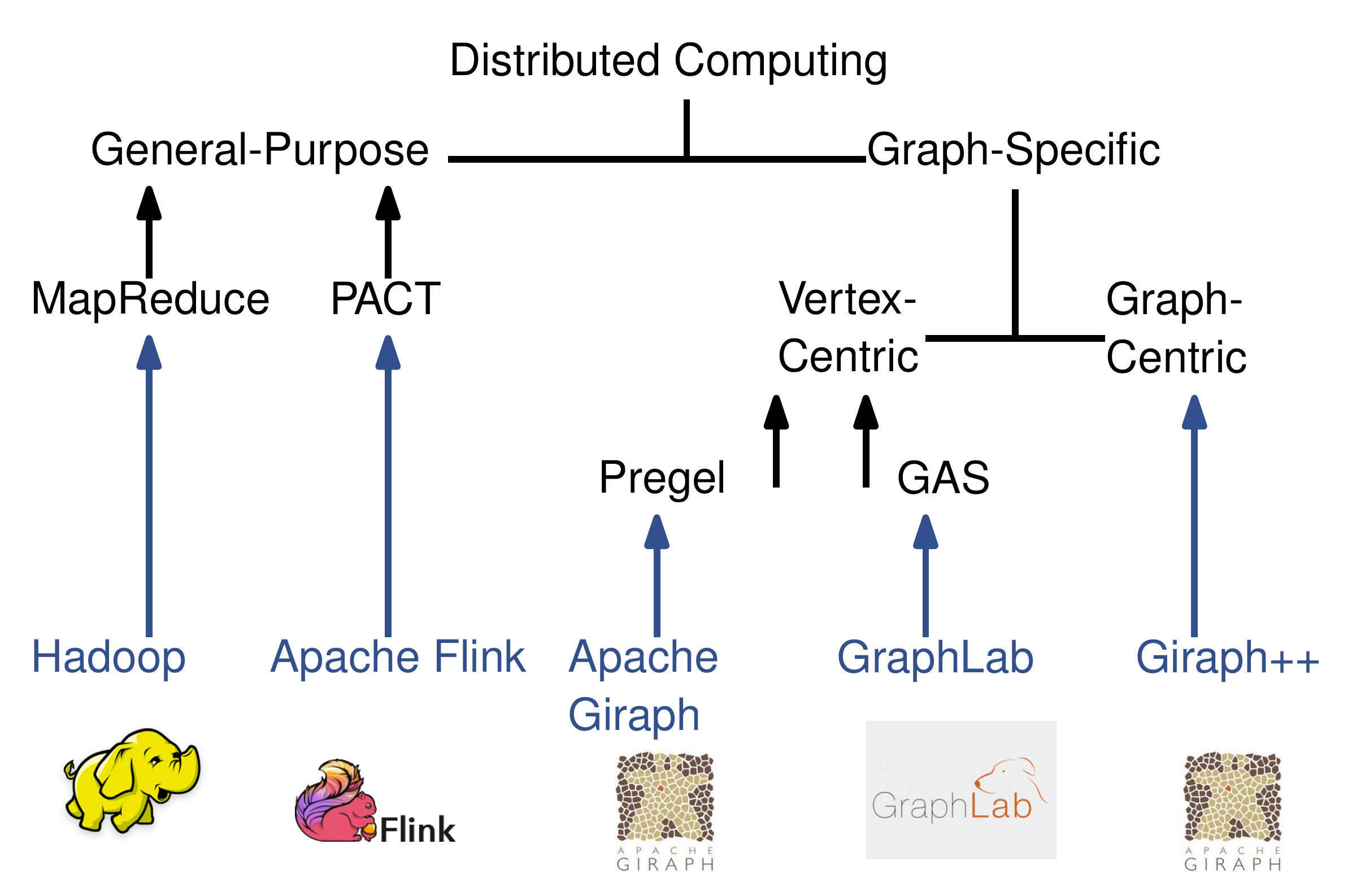}
\caption{Overview and classification of distributed programming models (black) and implementations (blue) considered in this paper.}
\label{fig:overview-models}
\end{center}
\end{figure}

Multiple programming models for distributed parallel computing have been introduced, some general-purpose (like MapReduce), others focus on the processing of large graphs (like Pregel) -- sometimes
called Big Graph frameworks to express their relation to Big Data. 
Among the software frameworks implementing these programming models, we consider Apache Giraph~\cite{webGiraph}, Apache Giraph++~\cite{giraph++}, GraphLab~\cite{webGraphLab}  and Apache Flink~\cite{webFlink} in our study.
An overview of programming models and their implementations is shown in Figure~\ref{fig:overview-models}.

Considering performance, the scalability and expressiveness enabled by such frameworks comes at the cost of overhead:
A recent experimental study~\cite{Satish:2014:NMG:2588555.2610518} compares implementations of graph algorithms based on frameworks (including GraphLab and Giraph) to ad hoc, hand-optimized code and observes a substantial performance gap.
Similar algorithms chosen for our study were also used recently by~\cite{McColl:2014:PEO:2567634.2567638}. They experience the aforementioned performance gap, too. However, they test
only on graphs generated with the synthetic R-MAT model, which lacks realism in several respects. 
Moreover, our focus is on distributed frameworks and we include some not used in their study.

Complex networks present certain inherent performance challenges for distributed systems:
Computation is data-driven in this case, i.e. it is largely determined by the graph structure. 
This makes it difficult to exploit opportunities for parallelism, because they depend heavily on the input. 
Real-world networks often have a skewed degree distribution, which adversely affects load balance.
In synchronous execution models, the few high-degree vertices will delay the whole system.
Performance depends also on how vertices are distributed among the processor nodes, i.e. the graph must be partitioned so as to reduce communication volume and to balance the work load -- a problem whose optimal solution is NP-hard to compute.
Given these challenges, performance experiments on real-world networks are crucial for choosing the right framework for the problem at hand.

\paragraph*{Contribution. \,}

We address the shortage of comparative studies and report performance results for several distributed frameworks.
Apache Giraph, GraphLab, Giraph++ and Apache Flink are described and compared regarding both theoretical and practical aspects.
The theoretical areas of comparison include the programming models and their suitability for expressing graph algorithms.
For an experimental comparison, several well-known graph algorithms were implemented using the frameworks.
With real-world complex networks (mainly from web applications) as input, the implementations were executed on a computer cluster to reveal how the frameworks compare in practice with regards to performance and memory footprint.
Additionally, we compare performance with implementations in a shared-memory parallel framework, NetworKit~\cite{staudt2014networkit}.
While the small-scale cluster we had available for this study is not enough to test out the true scalability of distributed frameworks, our results enable an informed choice on which framework to use for a given network analysis problem.
It becomes clear that, compared to shared-memory parallel solutions, distributed frameworks come with a significant overhead.
Consequently, opting for a distributed solution must be justified by the fact that the graphs to be processed cannot fit into main memory, which currently applies to graphs above the size range of several billion edges.

\section{Programming Models for Distributed Graph Processing}

\subsection{General-purpose Models}

\paragraph*{MapReduce and its Limitations. \,}

MapReduce is a programming model for general distributed computation~\cite{googleMapreduce}, and is often considered the de facto standard for this purpose~\cite{karloff2010model}.
A basic MapReduce program is defined by two user functions - a map and a reduce function.
Both functions receive and emit data as key-value pairs. 
A program is executed in three phases - a map, a shuffle and a reduce phase. 
Although MapReduce is able to express many common graph algorithms~\cite{LinTextProcessing}, it has limited practicality in this area:
Many graph analysis algorithms are iterative. 
Since the MapReduce model does not provide seamless support for iteration, this can only be reproduced by scheduling several consecutive  jobs, resulting in significant overhead.
Because each iteration needs the initial input graph data, the map function does not only emit an intermediate result for the input vertex, but also the vertex data itself, i.e. the vertex adjacency list and its current state. 
Consequently, graph data is sent over the network unnecessarily.
Workarounds like emitting the graph structure make MapReduce implementations harder to develop and understand.
We do not include MapReduce in the experimental study, since the models we discuss in the following offer more intuitive ways of expressing graph algorithms.

\paragraph*{PACT Model. \,}

The PACT ("Parallelization Contract") model~\cite{Battre2010} was proposed to extend MapReduce in both functionality and optimization opportunities.
It can be seen as a generalization of MapReduce that offers additional operators.
A PACT operator receives a user-defined function and an input data set containing tuples of predefined types. It applies the user-function to the input data in parallel and returns the output.
The operator is described by a) the \emph{Input Contract}, which defines how the user function can be applied to the input data set in parallel and b) the \emph{Output Contract}, which gives additional information on the data output. It can be optionally attached by the programmer and allows the PACT compiler to apply optimization techniques to the execution.
When writing a PACT program, the programmer uses an operator by defining its user function and the input data it operates on. The output of an operator can be used as the input of another, which creates a directed acyclic graph of operators (see Figure~\ref{fig:pact_graph_cc}). Unlike MapReduce, the PACT model does not force a specific order of the operators. 
The PACT compiler examines the structure of the resulting operator graph and infers an optimization strategy for the execution, \eg rearranging operators in the graph. A PACT program therefore has a declarative style, reminiscient of  the SQL language.
The PACT model is implemented by the Apache Flink framework~\cite{webFlink}.

\subsection{Vertex-centric Models}

Vertex-centric models express programs from a vertex perspective. These programs are executed iteratively for each vertex in the graph. The models are tailored to graph algorithms and, by incorporating vertices and edges, they include the data dependencies of graphs in the programming model. 

\paragraph*{Pregel. \,}

The Pregel model was introduced by Google in 2010 \cite{malewicz2010pregel}. It is based on the Bulk Synchronous Parallel model (BSP), which  is a general, iterative model for parallel computation \cite{valiant1990bridging}. An iteration consists of an algorithm that is executed on each processor, followed by a communication phase for data exchange between processors. Such an iteration is called a \emph{superstep}.
The execution is synchronous, thus a processor will only start executing a superstep if all other processors have finished with the previous one.
The Pregel model is implemented by the Apache Giraph framework~\cite{webGiraph}.

\paragraph*{Gather-Apply-Scatter (GAS). \,}

The GAS model was proposed as part of the PowerGraph abstraction~\cite{powergraph}, which is implemented in the GraphLab framework. 
GAS programs are iterative and vertex-centric like Pregel programs, but decompose an iteration into the gather, apply and scatter phases.
Accordingly, a GAS program is defined by a \texttt{gather}, \texttt{apply} and \texttt{scatter} and an optional \texttt{gather\_sum} function.
A further decomposition of the vertex program enables the framework to execute gather and scatter concurrently on several machines to balance the workload of high-degree vertices.
In the gather phase, a vertex can gather information from neighbors. The \texttt{gather} function is called for each edge of the vertex and returns the desired data.
The values returned are aggregated by the \texttt{gather\_sum} function.
The \texttt{apply} function is called once per vertex on the result of the gather aggregation.
Apply is the only phase where the vertex data can be changed.
Finally, the \texttt{scatter} function is called for each edge again. Scatter can be used to update edge data. 
The gather and scatter functions only have read access to vertex data. This allows them to be executed concurrently without the need for synchronization.
The GAS model is implemented by the GraphLab framework~\cite{webGraphLab}.

%

\subsection{Graph-centric Model}

\begin{figure}[h]
  \centering
  \includegraphics[width=.8\columnwidth]{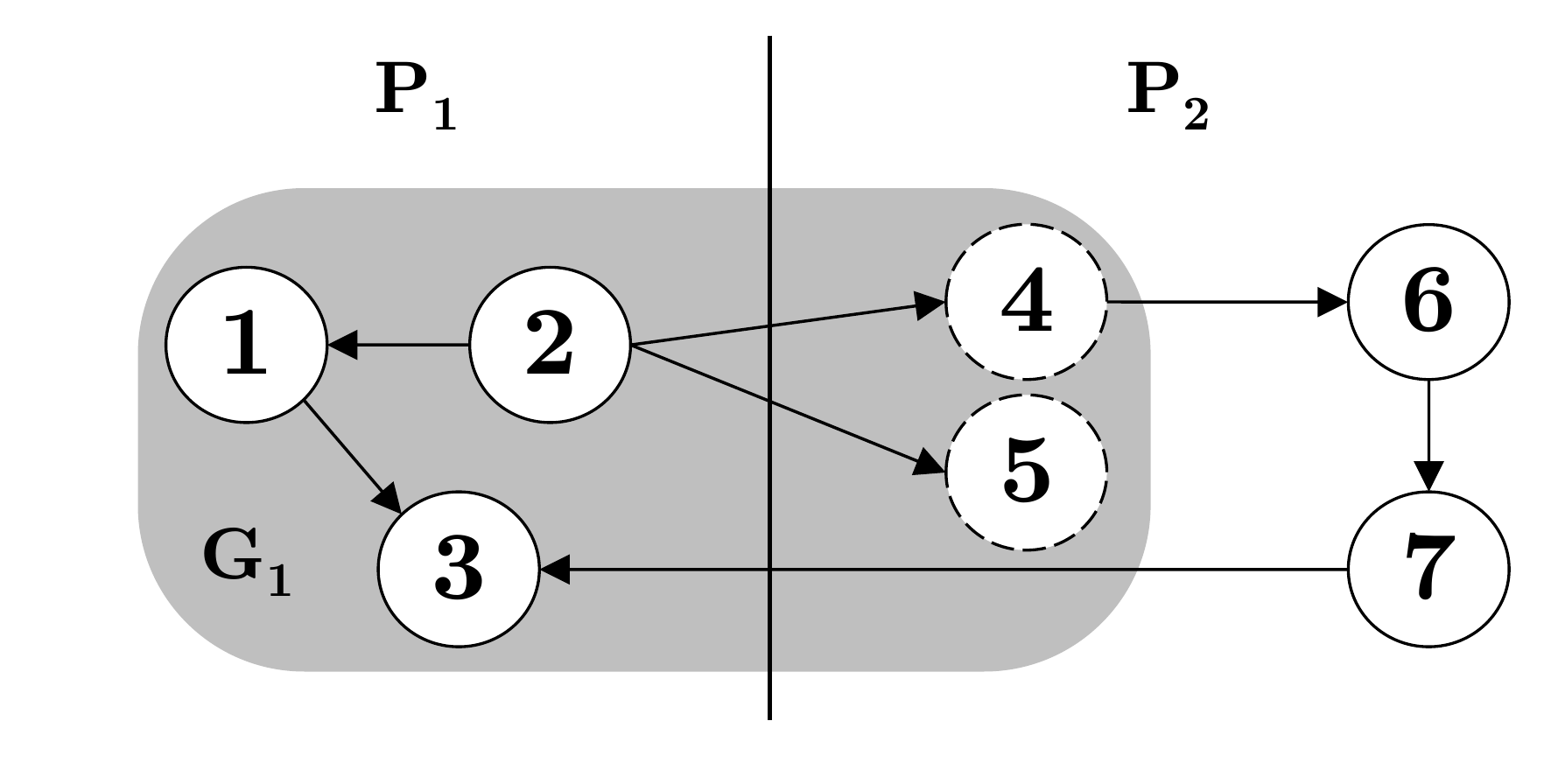}
  \caption[]{A graph with two partitions $P_1$ and $P_2$. The dashed vertices in $P_2$ are boundary vertices of $P_1$ and therefore part of the subgraph $G_1$.}
  \label{fig:internal_boundary}
\end{figure}

One of the reasons why vertex programs are so intuitive is that the abstraction hides details of the distributed execution, like graph partitioning.
However, this information could be used for algorithm-specific optimizations.
The graph-centric model~\cite{giraph++} is a lower-level abstraction which provides information on the partition structure to the programmer: A program is no longer expressed for one vertex, but for the whole block (or several blocks) of a partition.
Each vertex of the graph is part of exactly one block. The graph is divided into $k$ subgraphs $G_i$ (the blocks), each of which contains the vertices of a block $P_i$ as well as all vertices  adjacent to vertices of P$_i$. 
The vertices in P$_i$ are called \emph{internal vertices} of G$_i$, all others are called \emph{boundary vertices} (see Figure~\ref{fig:internal_boundary} for an example). 

A graph program is executed iteratively and synchronously, just like a Pregel program.
It is defined by a \texttt{compute} function, which can access the data of all vertices, both internal and boundary. Yet, only internal vertices can be changed. Updates to boundary vertices are only possible by sending a message to the corresponding internal vertex. The owner of the vertex receives incoming messages and can update it accordingly.

The lower-level graph-centric abstraction is useful in the following areas:
In vertex programs, data can be passed to neighbors only, so it can take many supersteps to propagate information across a graph. This is slow and could be accelerated for the vertices that are in the same partition. In the graph-centric model, messages need to be sent to boundary vertices only, as we can directly update vertex states of internal vertices. This reduces the amount of messages that are sent and processed. It may also reduce the number of supersteps: Information of one vertex is available to all vertices in the partition in the same superstep, which leads to faster convergence and therefore execution time for some problems.
The graph-centric model also allows the use of sequential algorithms as they can be easily applied to the subgraph represented by the partition. 
Graph-centric programs are executed synchronously, as a synchronization barrier waits for the last computation to finish after each superstep. However, local asynchrony is possible, as updates made to vertex data are available within the computation of one partition immediately.
The graph-centric model is implemented by the Giraph++~\cite{giraph++} framework.

\section{Frameworks}

This section introduces the software frameworks that implement the aforementioned programming models. All of them share some common implementation principles, which we discuss briefly before the specifics of each framework: 
One is the use of a \textit{distributed file system}, which provides the means to store data sets that are too large for a single machine and enable data locality and fault-tolerance. Files are split up into blocks which can distributed across a cluster. The file system coordinates how clients access and write to files. All frameworks presented support the \textit{Hadoop Distributed File System (HDFS)}.

The frameworks of message-based programming models also implement \textit{message combiners} as follows: In many cases, the message receiver is not interested in each individual message value, but some kind of aggregation of the messages. Therefore, messages can be combined on the sender node to reduce network traffic. The principle of combiners can be used in vertex- and graph-centric models by implementing a \texttt{combine} function which is called by the framework on messages with the same receiver.

\subsection{Apache Flink}

Apache Flink~\cite{webFlink} (previously called Stratosphere~\cite{webStratosphere}) is an open-source framework written in Java. 
It implements the PACT programming model. PACT was put forward by this project with the goal of extending the MapReduce model with new operators and a flexible compiler able to infer optimization strategies from descriptive programs. 
Apache Flink offers a Java and a Scala API to write PACT programs. %

\subsection{Apache Giraph}

Apache Giraph~\cite{webGiraph} is an open-source implementation of the Pregel programming model. 
It is written and implemented in Java and uses Apache Hadoop, a MapReduce implementation, for the execution of vertex programs.
Facebook uses Apache Giraph in practice for the analysis of its social graph and is an active contributor to the project~\cite{facebooktrillion}. 

\subsection{GraphLab}

The GraphLab~\cite{webGraphLab} package was initially intended to be used on shared memory systems, but was extended to support distributed execution~\cite{graphlabPreGAS}. Programs utilizing GraphLab are written in C++; Python bindings are also available.
With GraphLab 2, support for the GAS model was introduced, which was proposed as part of the PowerGraph abstraction \cite{powergraph}. 

\subsection{Giraph++}

The graph-centric model originates from a paper by Tian et. al. \cite{giraph++}, who also put forward Giraph++, an implementation of this model developed on top of Apache Giraph. Currently Giraph++ is not available as a separate framework, but as a patch for Apache Giraph which adds support for the graph-centric model to Giraph~\cite{giraph++patch}. It is proposed to become part of the Apache Giraph project.

\section{Network Analysis Algorithms}
\label{sec:algorithms}

For the experimental study, we select four common and typical network analysis problems, Connected Components, Community Detection, PageRank, and Clustering Coefficients.
The former two call for a decomposition of the entire network into cohesive parts, and we opt for algorithms based on label propagation.
Thus, both lend themselves to vertex-centric implementations.
The latter two yield a ranking of vertices by structural relevance, PageRank through a simple iterative rule, Clustering Coefficients based on the somewhat more intricate counting of triangles.
We consider the program flow of these algorithms representative for many algorithmic kernels used for network analysis. Moreover, their
algorithmic structure is able to highlight challenges and benefits of distributed processing.

\subsection{Connected Components via Label Propagation}
Many real-world networks have several connected components, but one of them usually dominates by far in size~\cite{Newman2010}.
For network analytic purposes it is interesting to determine whether a vertex belongs to this giant component or not.
Thus, subdividing the network into connected components is a common analysis task and can be implemented as attaching a component label to each vertex.
The label propagation algorithm used in this study starts with each vertex in its own component and communicates its component ID to all neighbors.
A vertex then applies the lowest component ID to itself.
If the label has changed, the vertex will notify all neighbors with the new component ID.

\begin{figure}[h]
  \centering
  \includegraphics[width=\columnwidth]{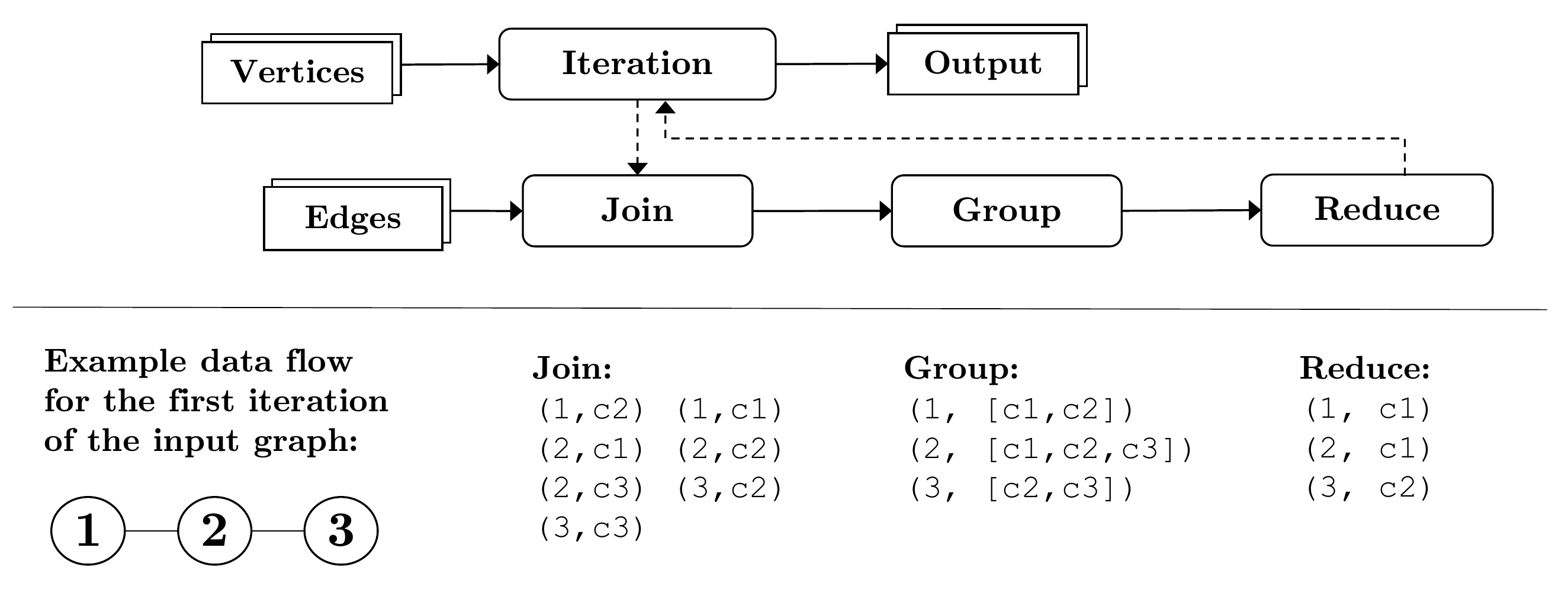}
  \caption[PACT operator graph of Connected Components]{PACT operator graph of the Connected Components example}
  \label{fig:pact_graph_cc}
\end{figure}

Figure \ref{fig:pact_graph_cc} illustrates the principle of the algorithm in PACT: The algorithm works on tuples of vertex IDs and component labels, whereas the edge information is used to create these tuples. 

\begin{figure}[h]
  \centering
  \includegraphics[width=.7\columnwidth]{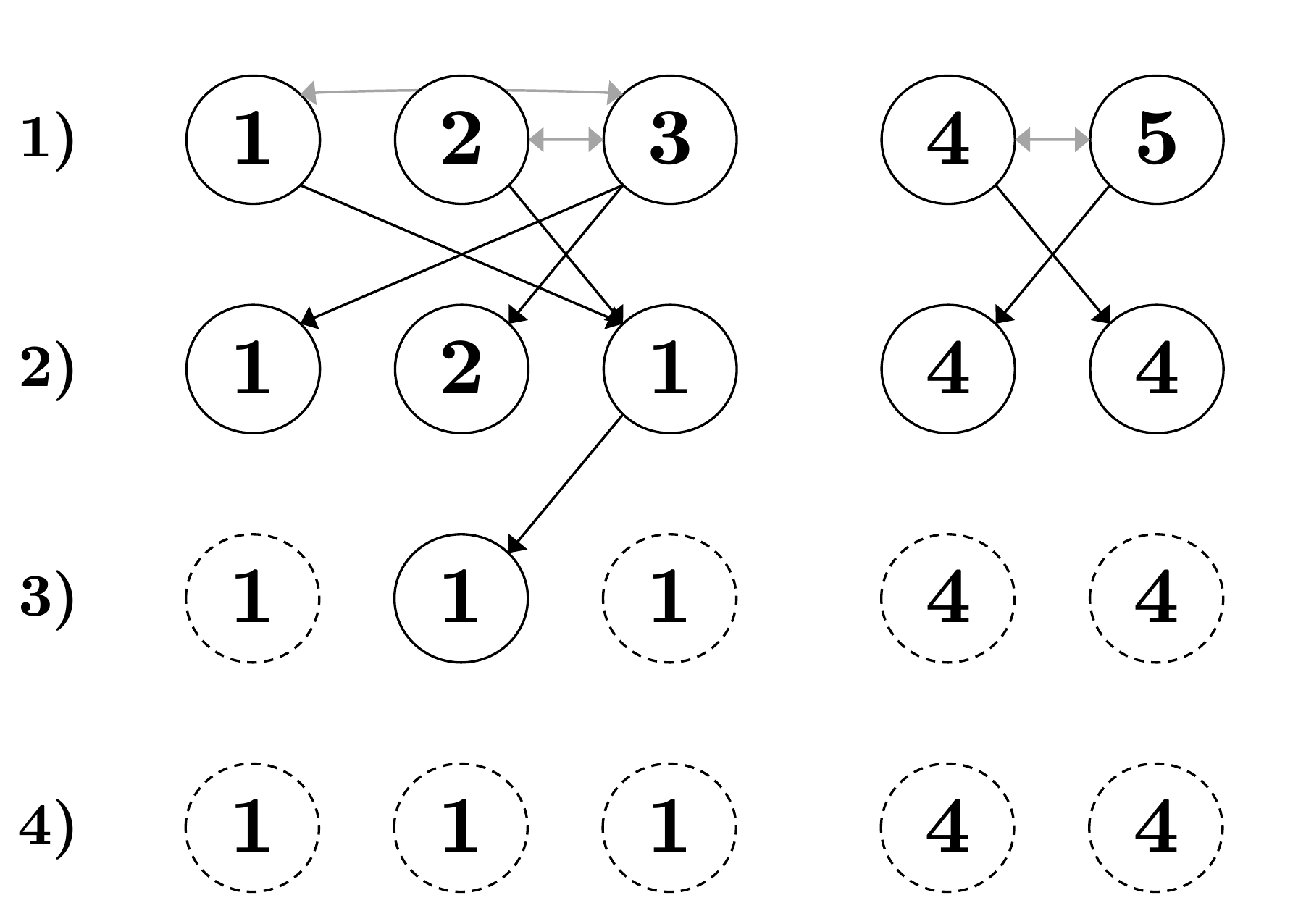}
  \caption[]{Example computation of connected components in the vertex-centric Pregel model. The two connected components
are separated by a larger gap in the middle. Gray arrows represent graph edges, black arrows represent message passing, inactive vertices are indicated by a dotted line.}
  \label{fig:pregel_example_cc}
\end{figure}

The implementations in Pregel and GAS can be seen in Algorithms~\ref{alg:pregel_cc} and~\ref{alg:gas_cc}: After the initialization with the vertex ID as component label, the vertices distribute their component label along their edges to neighbors and in turn apply the lowest component label from all incoming labels. Figure~\ref{fig:pregel_example_cc} shows an example run of the algorithm in the Pregel model.
\input{codes/ch02_pregel_cc}
\input{codes/ch02_gas_cc}	

\begin{figure}[h]
  \centering
  \includegraphics[width=.8\columnwidth]{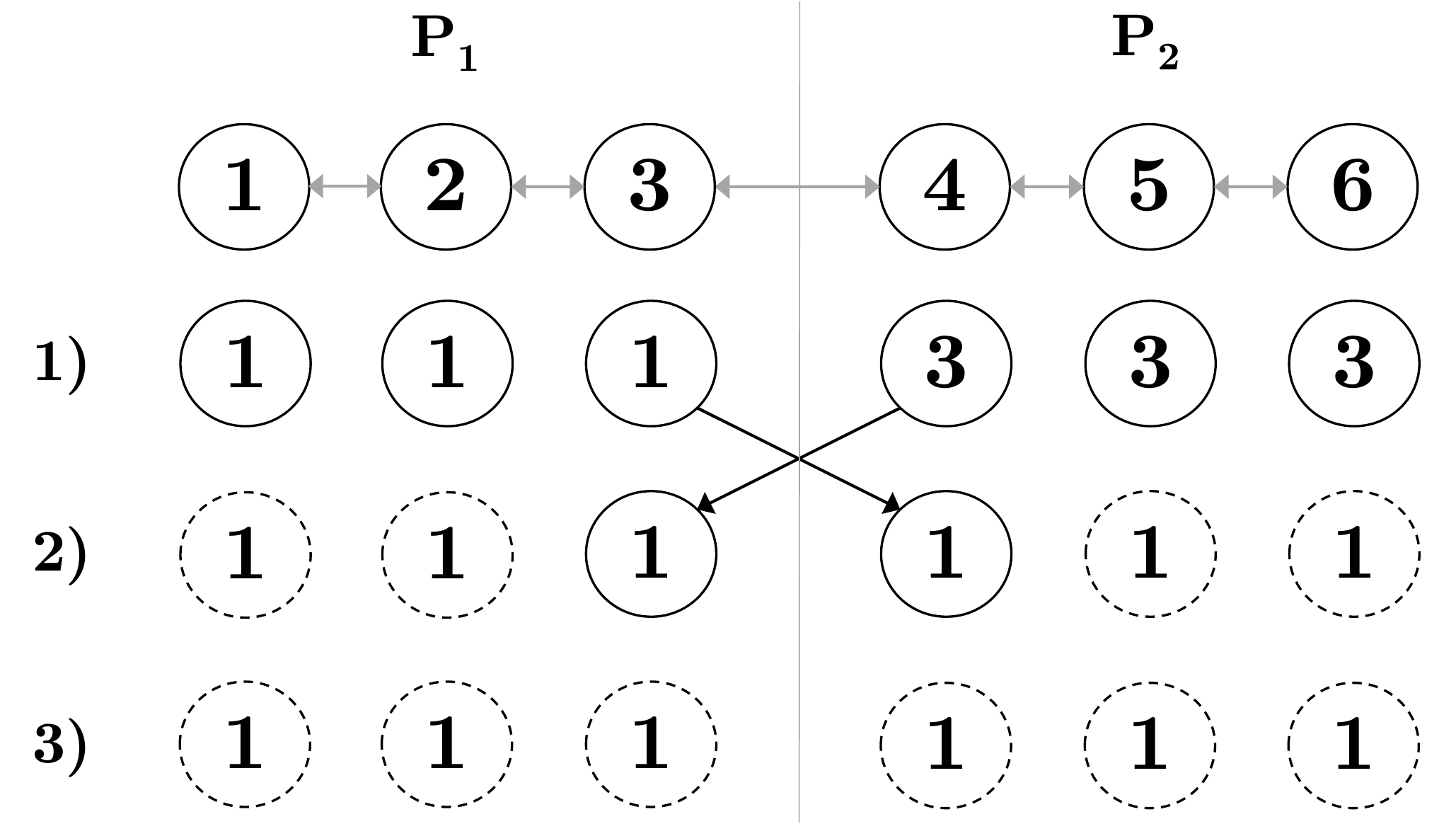}
  \caption[]{Example computation of connected components in the graph-centric model of Giraph++: For each boundary vertex, a message with the component label is sent to the partition with the corresponding internal vertex.}
  \label{fig:giraph++_example_cc}
\end{figure}

In the graph-centric model, the algorithm does the following: In the first superstep, a sequential algorithm is used to find all connected components in one partition. For each boundary vertex, a message with the newly found component label is sent to the partition with the corresponding internal vertex. If any of the incoming messages have a smaller label for a vertex, then this vertex and all other vertices within the same component and partition will be updated at once. Figure~\ref{fig:giraph++_example_cc} shows an example run of the computation.

\subsection{Community Detection via Label Propagation}
Many real-word networks exhibit a \emph{community structure}, \ie they can be decomposed into internally dense and externally sparse subgraphs named communities.
Among heuristics for detecting communities, a label propagation algorithm~\cite{raghavan2007near} is among the fastest and easiest to parallelize:
Each vertex is initialized with a unique community label, and vertices propagate their label to all neighbors in each iteration and update it with the most frequent of the received labels. Ties are broken systematically, \eg uniformly at random or by preferring
the minimum label.
The algorithm can be executed synchronously or asynchronously. As the synchronous execution can lead to an oscillation of labels \cite{raghavan2007near}, the asynchronous execution is preferred.
When adapted suitably, the general idea of label propagation is also useful for partitioning a network into equally sized blocks
with few edges running between different blocks~\cite{MeyerhenkeSS14partitioning,DBLP:conf/bigdataconf/SlotaMR14}.
Such results are potentially useful for computing partitions in the graph-centric model.

In PACT, the algorithm is expressed as follows: The label of a vertex is initialized with the vertex ID of a random neighbor. A bulk iteration is performed on the resulting data set. In each iteration, a \texttt{join} operator is used to get tuples (\texttt{vertexId}, \texttt{neighborLabel}) for every vertex. A \texttt{group} operator on both vertex and label followed by a \texttt{sum} aggregator counts the labels for each vertex. This data set is then \texttt{group}ed by the vertex ID, a \texttt{reduce} operator finds the most frequent labels and emits the result.

The label propagation algorithm can be naturally expressed in the vertex-centric model of Pregel.
In the algorithm's initialization superstep, an optimization is possible since in Pregel each vertex can access
its neighbors' IDs. Since these IDs are equal to the initial label value, there cannot be a most frequent one.
Thus, the label of each vertex is initialized with the label of a randomly selected neighbor.
This label will be sent to all neighbors, which select the most frequent of the incoming labels.
If the new label differs from the previous one, a local \texttt{converged} flag is set to false. 
Global aggregation of this flag is used to check for termination.

In GAS the algorithm works similarly, as \texttt{gather} and \texttt{gather\_sum} determine the label frequency, \texttt{apply} applies the new label and \texttt{scatter} distributes it among neighbors.

The graph-centric implementation is again similar, with the major difference that within a partition the labels can be updated directly, because messages have to be exchanged for boundary vertices only. Since the label updates are not applied globally in a synchronized way, label oscillation is less likely to occur.

\subsection{PageRank}
PageRank is a well-known centrality algorithm, enabling a ranking of vertices by their structural importance -- and taking the importance of
neighbors into account.
It was originally developed to rank websites in the web graph~\cite{Brin98theanatomy}, but due to Google's success
it has experienced much wider attention
and applications~\cite{Newman2010}. The PageRank corresponds to the dominant eigenvector of a transition matrix modeling
a random surfer. In practice one often approximates a vertex score $P(v)$ by using the power iteration algorithm, \ie by 
computing the iteration $P_{i+1}(v) = \alpha r(v) + (1 - \alpha) \cdot \sum_{\{u,v\} \in E} \frac{P_i(u)}{deg(u)}$,
where $\alpha$ is a dampening factor that models a random jump to an arbitrary webpage (as opposed to following a link). 
In the following we make the usual assumption of $r(v) = 1$ for all $v \in V$ and call $P(v)$ the \emph{score} of vertex $v$.

The PACT implementation works on different sets of tuples: 1) vertex ID and score and 2) vertex ID and a list of neighboring vertices. A bulk iteration for the vertices is then defined by a \texttt{join} over the neighbor tuple and the corresponding scores. Afterwards, a \texttt{group} by vertex ID and the aggregation by computing the sum follows.

In GAS, the PageRank algorithm is easily expressed as follows (also see Algorithm~\ref{alg:pagerank_gas}): \texttt{gather} is called for all incoming edges and returns the neighbor vertex score divided by its outdegree. 
\texttt{apply} receives the sum of these values, which is used to calculate the new score and the \texttt{delta} value. If \texttt{delta} is below the tolerance, the \texttt{converged} flag is set to true. \texttt{scatter}, which is called on outgoing edges, signals neighbor vertices, but only if the vertex score has not converged.
In Pregel, the termination is handled differently as all vertices remain active as long as the global termination criterion is not met
(Algorithm~\ref{alg:pagerank_pregel}).

\input{codes/ch04_pagerank_gas}

\input{codes/ch04_pagerank_pregel}

The graph-centric implementation is again similar to the vertex-centric models, in this case Pregel. However, in Giraph++ the support for local asynchrony is used: Values from vertices within the partition can be accessed directly while for boundary vertices messages need to be exchanged. Also, here we can make use of a minor optimization~\cite{accumulativePagerank} that was shown to converge faster than the traditional iterative approach.

\subsection{Clustering Coefficients}

Transitivity is a network analytic concept to express if two neighbors of a vertex are likely to be neighbors as well or not.
In social networks, for example, two friends of the same person are more likely to be friends as well than a random pair
of vertices~\cite{Newman2010}.
Thus, as one formalization of the transitivity, we can use the \emph{average local clustering coefficient} $C(G)$ for an undirected graph $G$ to examine its local connectivity. This particular measure is defined as the mean over the local clustering coefficient over all vertices, \ie $C(G) = \frac{1}{|V'|} \sum_{v \in V'} \frac{\delta(v)}{\tau(v)}$, where $V' := \{v \in V ~|~ \deg(v) \geq 2)$, 
$\delta(v)$ is the number of triangles $v$ belongs to, and $\tau(v) := {\deg(v) \choose 2}$ (see \eg~\cite{schank2004approximating}).
This is different from the \emph{global clustering coefficient}, which 
contrasts the numbers of closed triangles and of connected triplets in a graph (times a constant factor).

All of the described measures can be computed with the same algorithmic idea, as the basic operation is counting triangles. Since the exact algorithm is computationally expensive for large complex
networks, we are also interested in an efficient approximation algorithm. 
\cite{schank2004approximating} proposed a sequential random sampling algorithm to approximate the clustering coefficient with a (probabilistic) additive error bound. For the \emph{average local clustering coefficient}, it randomly selects a vertex $v$, and two different neighbors of $v$, $u$ and $w$.  If $u$ and $w$ are connected, a counter is incremented. This process is repeated for a fixed number of samples. The approximation value of the \emph{average local clustering coefficient} is the number of links found divided by the number of samples. For the \emph{global clustering coefficient} the samples will be chosen with a probability proportional to $deg(v) \cdot (deg(v)-1)$ for a vertex $v$, where $deg(v)$ denotes the degree of $v$.

In PACT, to count the relevant edges, two \texttt{join} operations are necessary. The first is carried out on the edges with themselves resulting in a set of neighbors of neighbors of a vertex. The second \texttt{join} then joins the resulting set with the edges again with respect to a vertex. The rest of the algorithm is straightforward.

The Pregel implementation (Algorithm~\ref{alg:clustering_pregel}) consists of two supersteps where the first one sends the edge information to incident vertices while the second one computes the local clustering coefficient and sets the aggregators for the global values. 

The GAS implementation (Algorithm~\ref{alg:clustering_gas}) is similar as it uses one step to calculate the common neighbors of two vertices and uses this information in the second step to do the final computation. Aggregation of the vertices' values is then done 
using mapReduce functions.
The graph-centric model does not yield any advantages over the vertex-centric models in this application, so the implementation basically follows the Pregel implementation. Again, the main difference is that the computation for edges in a partition can be done directly, whereas explicit message exchange is necessary for boundary vertices.

The implementation of the approximation algorithm principally follows the implementation of the exact algorithm with a few exceptions. Depending on whether the global or the average local clustering coefficient is computed, the probability for each vertex needs to be set. Then, instead of exchanging the whole neighborhood, each vertex will only draw a fixed number of vertex pairs and send it to these vertices, e.g.\ during a superstep in Pregel. Another superstep is then used to intersect the neighborhoods and the incoming messages and for each match the link counter will be incremented. Afterwards, the usual global aggregation will compute the final values.

\input{codes/ch04_clustering_pregel}

\input{codes/ch04_clustering_gas}

\subsection{Discussion}
In summary, we confirm that graph-specific frameworks (Giraph, Giraph++, GraphLab) offer a more concise and intuitive way to express graph algorithms.
Not surprisingly, vertex-centric programming models are well suited for vertex-centric algorithmic techniques as in label propagation or PageRank's iterative method.

\section{Experimental Comparison}

In the following we present a comparative experimental study on the performance of four distributed and one single-machine framework.

\subsection{Experimental Setup}

For the experiments a rather modest computer cluster of 8 nodes was used. Each system has 24GB RAM and runs an Intel\textsuperscript{\textregistered} Xeon\textsuperscript{\textregistered} X5355 CPU with 2.66 GHz and 8 cores on two processors. 
The framework versions are GraphLab 2.2, Apache Giraph 1.1.0, Apache Flink 0.6, and NetworKit 3.3. (The NetworKit version has
been patched with a fix for an earlier performance bug in the PageRank implementation.)
Apache Giraph was deployed in a Apache Hadoop MapReduce environment, using Hadoop 0.20.203.0 and Apache ZooKeeper 3.4.6. 
For scaling experiments, the algorithms were run on 1, 2, 4 and 8 cluster nodes. NetworKit was run on a single node of the same type.

\paragraph*{Fault tolerance.\,}
Apache Giraph is the only one of the examined frameworks which supports the recovery in case of a failure during the algorithm's execution. It uses a checkpoint mechanism which saves the current computation state of the algorithm execution to the distributed file system. Workers regularly send heartbeat messages to the Master to indicate they are executing, thus the absence of a heartbeat message indicates a failure. In this case, the master starts up new workers replacing the failed ones. Then, all workers load the last checkpoint from the file system and resume the computation.
The checkpointing mechanism was tested in experiments for this particular purpose by killing off one worker process. The master detects the missing worker immediately, and computation resumes without errors. However, the overhead for saving checkpoints regularly slows down the performance.
Although GraphLab has a similar checkpoint mechanism, it does not automatically recover in the case of a failure. GraphLab periodically saves a binary serialization snapshot of the graph. The algorithm has to be restarted manually using the snapshot as the input graph.

\paragraph*{A Shared-Memory Framework for Comparison.\,}

NetworKit~\cite{staudt2014networkit} is a tool suite of algorithms and data structures for the analysis of large complex networks, implemented in C++, OpenMP in the backend and in Python 
for a user-friendly frontend. 
NetworKit runs on a single computer, its OpenMP implementations exploit multicore capabilities.
It is based at Karlsruhe Institute of Technology and developed with international contributors under an open-source model.\footnote{\url{http://networkit.iti.kit.edu}}
The aim of the package is to provide efficient, often parallelized graph algorithms and lean data structures for network analysis.

For this study, we let efficient shared-memory parallel implementations of network analysis algorithms from NetworKit compete with implementations in distributed frameworks.
The aim of the comparison is to see when it is worth to move to distributed systems.

\newcommand{\graFlickrEdges}{\texttt{flickr-edges}\xspace}
\newcommand{\graFlickrLinks}{\texttt{flickr-links}\xspace}
\newcommand{\graOrkut}{\texttt{orkut-links}\xspace}
\newcommand{\graLive}{\texttt{livejournal-links}\xspace}
\newcommand{\graLiveU}{\texttt{livejournal-links-u}\xspace}
\newcommand{\graLiveD}{\texttt{livejournal-links-d}\xspace}
\newcommand{\graUk}{\texttt{uk-2002}\xspace}
\newcommand{\graWikiLinksEn}{\texttt{wikipedia-links-en}\xspace}
\newcommand{\graTwitter}{\texttt{twitter}\xspace}
\newcommand{\graTwitterL}{\texttt{twitter-l}\xspace}

\paragraph*{Test data sets.\,}
Real world graph data sets used for the experiments are listed in Table~\ref{tab:graphs}. The \graUk data set \cite{BoVWFI} originates from a web crawl of the .uk domain. The remaining data sets were taken from the KONECT graph collection \cite{konect}. The Wikipedia data set consists of linked encyclopedia pages of the English Wikipedia.  The \graOrkut, \graLive, \graFlickrLinks graphs represent the users and friendships of the respective social networks. \graTwitter \cite{twitter1} and \graTwitterL \cite{twitter2} are snapshots of the follower graph of the social network Twitter. \graTwitterL  is the largest data set used in the experiments with almost two billion edges, taking up over 36 GB of disk space when represented in an edge list file. The relatively small \graFlickrEdges graph is based on images of the Flickr website, adjacent images share common meta data.
Additionally, synthetic graphs (scale-free graphs generated using NetworKit's Dorogovtsev-Mendes generator) are used for weak scaling experiments.
These synthetic data sets contain 16 million edges per compute node, i.e. 16, 32, 64 and 128 million edges for the experiments on a single, 2, 4 and 8 compute node(s).

\begin{table}
\begin{center}
\begin{scriptsize}
    \begin{tabular}{| l | r | r | r |}
    \hline
        
     name (directed) & n & m & size\\ \hline
    \graTwitterL (1)& 52M & 1,963M & 36.16\\ \hline
    \graTwitter (1)& 41M & 1,468M & 22.35\\ \hline
    \graWikiLinksEn (1)& 27M & 601M & 9.48\\ \hline
  	\graUk (1)& 18M & 261M & 8.26\\ \hline
  	\graOrkut (1)& 3M & 117M & 1.65\\ \hline
    \graLiveD (0)& 4.8M & 68M & 0.99\\ \hline
    \graLiveU (1)& 5M	& 49M & 0.67\\ \hline
    \graFlickrLinks (0)& 1.7M & 15M & 0.18\\ \hline
    \graFlickrEdges (0)& 105k & 2.3M & 0.02\\ \hline
    \end{tabular}
    \end{scriptsize}
\end{center}
\caption{Properties of the networks (with $n$ vertices and $m$ edges) used in the experiments. Size denotes the size of the graph stored in edge list format in GB.}
\label{tab:graphs}
\end{table}

\subsection{Results}

\paragraph*{Connected Components via Label Propagation.\,}

\begin{figure}[h]
  \centering
  \includegraphics[trim=0 385 100 0,clip, width=.5\textwidth]{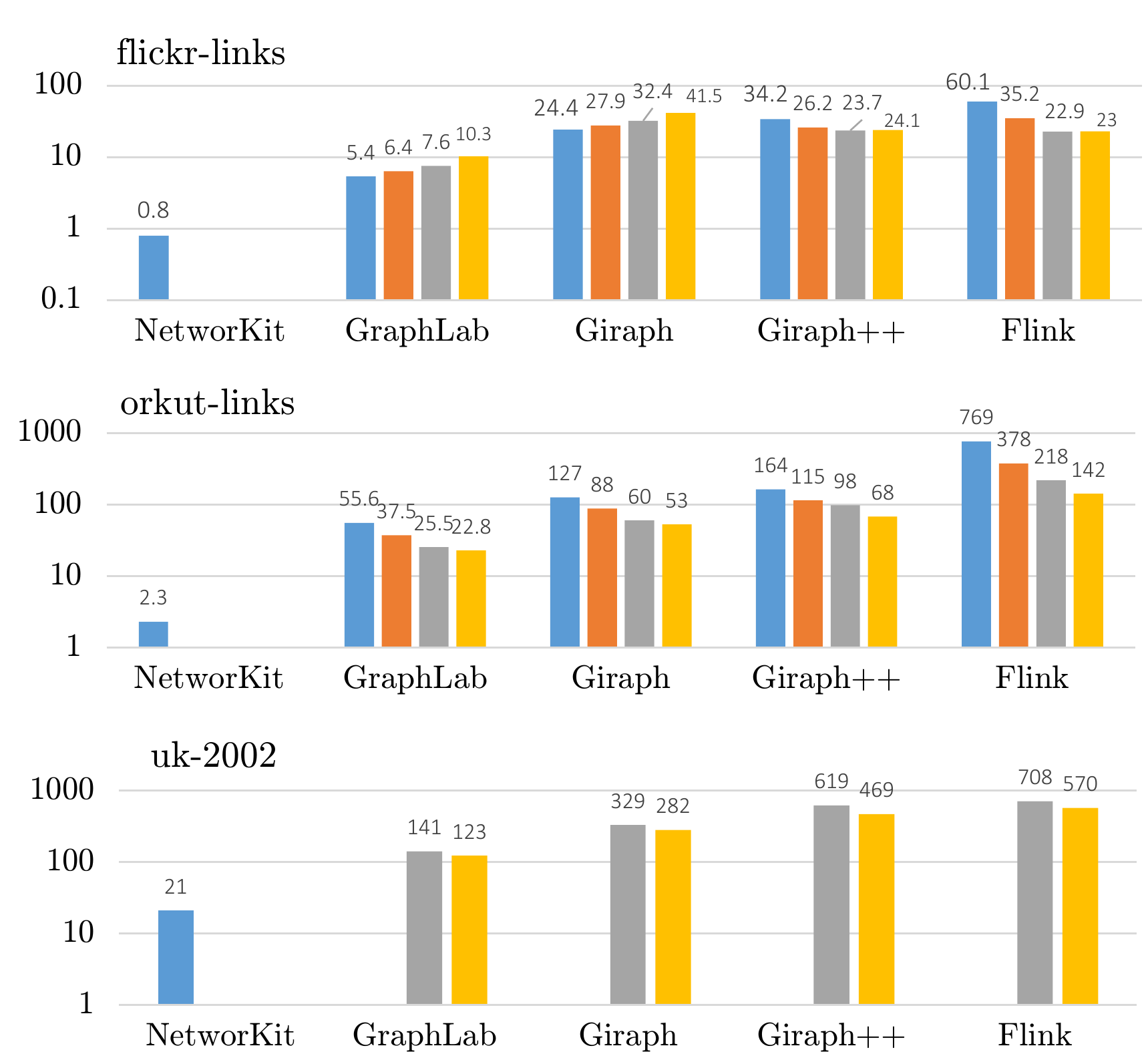}
   \includegraphics[trim=0 190 100 630,clip, width=.5\textwidth]{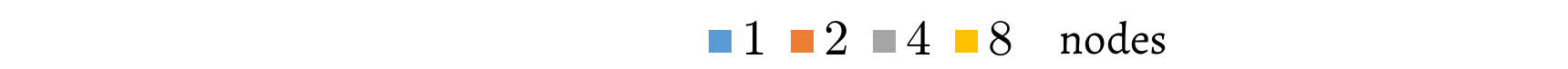}
    \caption[Connected Components running times]{Running times [in seconds] of Connected Components.}
    \label{fig:perf_cc}
  \centering
\end{figure}

As Figure~\ref{fig:perf_cc} shows, the shared-memory parallel implementation in NetworKit is much faster than the distributed frameworks. Out of the latter, GraphLab performs best, followed by Apache Giraph. Giraph++ cannot benefit from graph-centric optimizations and although it reduces the number of supersteps, it performs worse. Apache Flink is much slower than the other frameworks on a single node (over 13x slower than GraphLab and 6x slower than Giraph for the \graOrkut graph). It catches up in the cluster setup due to higher speedup rates, but remains slower than GraphLab and Giraph.
For smaller graphs such as the \graFlickrLinks graph, GraphLab and Apache Giraph running times increase with adding additional compute nodes due to overhead, while Giraph++ and Apache Flink do achieve a speedup. 
The connected components on \graUk can only be computed with at least 4 nodes by any of the distributed frameworks. 
The billion-edge \texttt{twitter} network can only be processed with GraphLab and Giraph (besides NetworKit), and the two distributed
frameworks require at least 8 nodes.
Figure~\ref{fig:concomp_weak} shows that only the Giraph implementation has a stable weak scaling behavior, and that GraphLab has the largest relative increase, with the time quadrupling from the single node to the eight node execution.

\input{img/plots/concomp_weak.tex}
\begin{figure}[h]
  \centering
  \includegraphics[trim=0 780 242 0,clip, width=.5\textwidth]{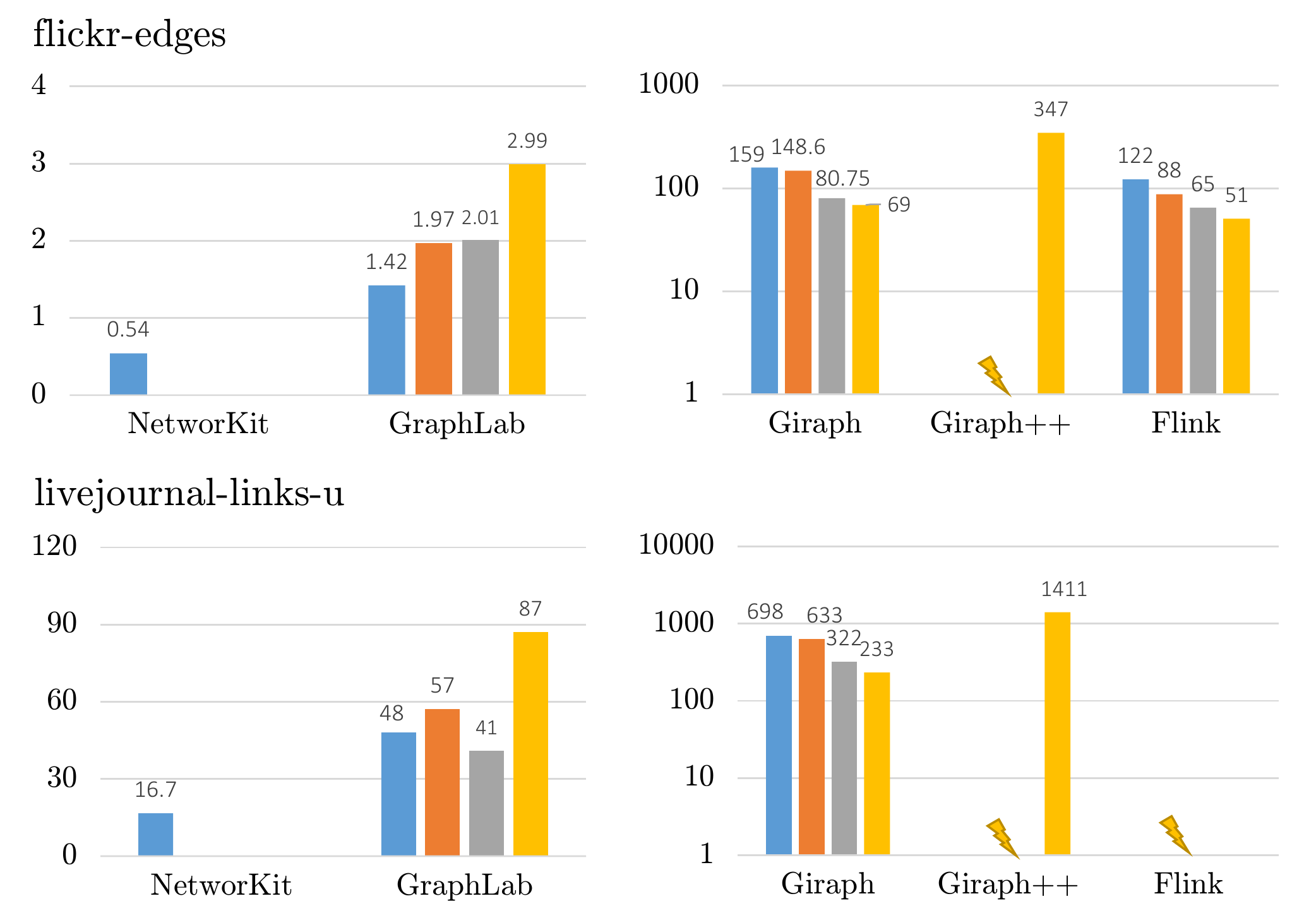}
  \includegraphics[trim=0 190 100 630,clip, width=.5\textwidth]{img/plots/cc_times_legend.pdf}
    \caption[Community Detection: running times]{Running times [in seconds] of Community Detection on 1, 2, 4 and 8 nodes. The flash symbol indicates that a graph is too large to be processed with the respective tool using the given number of nodes.}
    \label{fig:perf_cdlp}
  \centering
\end{figure}

\paragraph*{Community Detection via Label Propagation.\,}

Out of the distributed frameworks, GraphLab has a clear advantage over the other frameworks with its asynchronous model, which is evident in the running times (Figure \ref{fig:perf_cdlp}).
 NetworKit is again much faster.
 For \graFlickrEdges and \graLiveU, GraphLab gets slower when more computing nodes are added. 
 Flink can only handle the \graFlickrEdges data set. Giraph++ computes data sets up to the \graOrkut graph, but is very slow (6x the execution time of Giraph for \graLiveU). Both GraphLab and Giraph process graphs up to the size of \graWikiLinksEn on a cluster of eight nodes.
  The synchronous execution of this algorithm can cause convergence problems that prevented some implementations from terminating. Label oscillations can occur as two neighboring vertices join the community of the other vertex and because the communication happens synchronized after local computations are done, they effectively exchange their labels. This is a disadvantage of frameworks only supporting synchronous execution (Giraph, Giraph++ and Flink) for this type of algorithms.

\paragraph*{PageRank.\,}

With the exception of NetworKit's implementation, the PageRank implementations tested run for a fixed number of iterations.
PageRank was executed on the directed test graphs. The \graLiveD graph could be processed by all frameworks in every setting. \graUk and \graWikiLinksEn could be processed by the distributed frameworks with 4 and 8 nodes only, \graTwitter only in the 8 node setting. The largest graph, \graTwitterL, could only be processed by Apache Giraph. GraphLab stopped with memory allocation errors and Flink aborted the computation. 

\begin{figure}[h]
  \centering
\includegraphics[trim=0 445 598 0,clip, width=.5\textwidth]{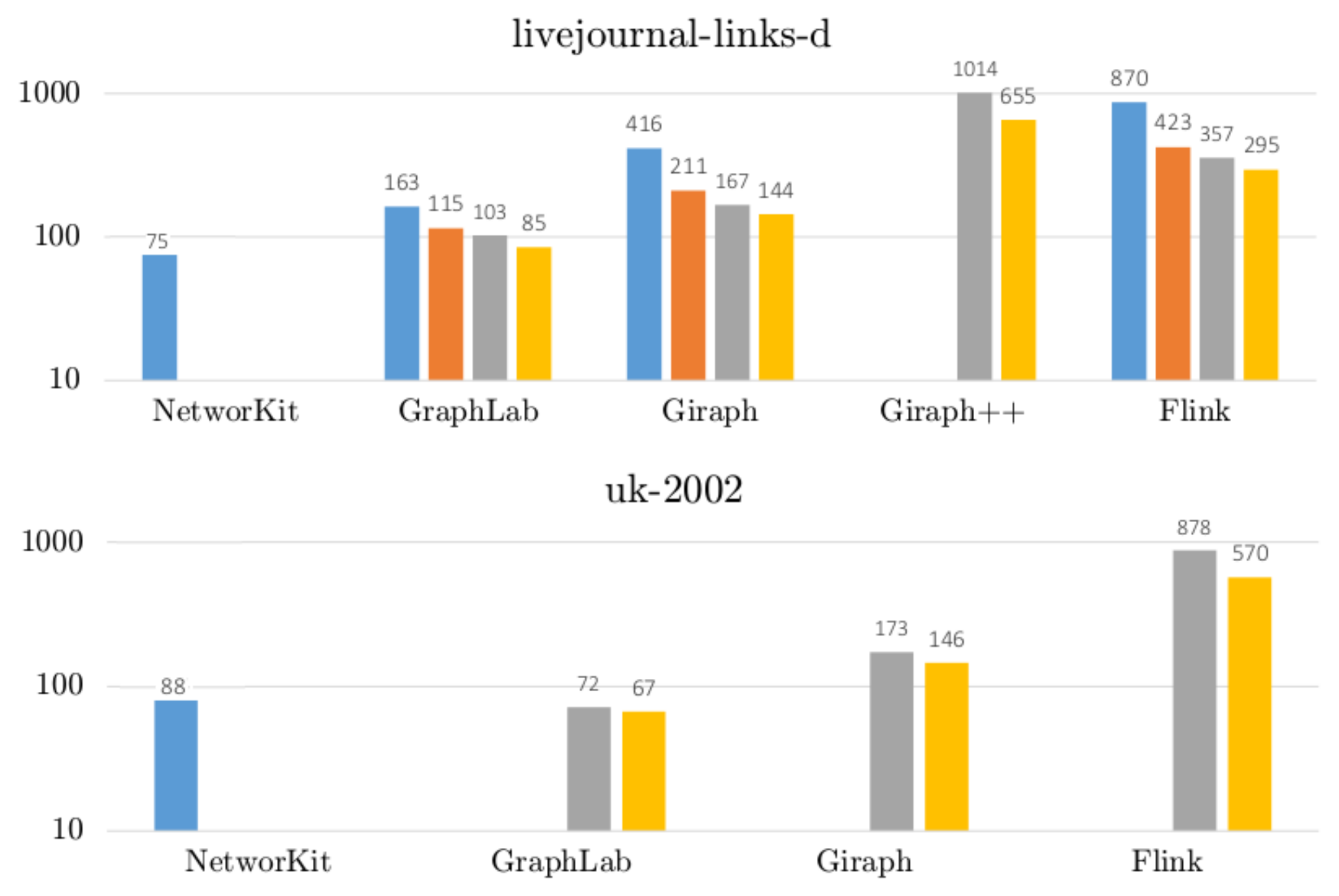}
  \includegraphics[trim=0 190 100 630,clip, width=.5\textwidth]{img/plots/cc_times_legend.pdf}
    \caption[PageRank: running times]{Running times [in seconds] of PageRank.}
    \label{fig:perf_pr}
  \centering
\end{figure}

GraphLab and NetworKit achieve the best performance for PageRank (see Figure~\ref{fig:perf_pr}): Apache Giraph is slower than GraphLab by a factor of 1.6 to 2.5 for the smaller \graLiveD, and 2.8 to 3 for the larger \graWikiLinksEn and \graTwitter. The differences to Apache Flink are even larger, the latter's running time is up to 12 times (\graUk) higher than that of GraphLab.
The Giraph++ algorithm exhibits poor performance.
GraphLab is sometimes faster than NetworKit due to the fact that NetworKit iterates longer until a higher accuracy is achieved.

%

\paragraph*{Clustering Coefficients.\,}
\begin{figure}[h]
  \centering
  \includegraphics[trim=0 85 0 105,clip, width=.5\textwidth]{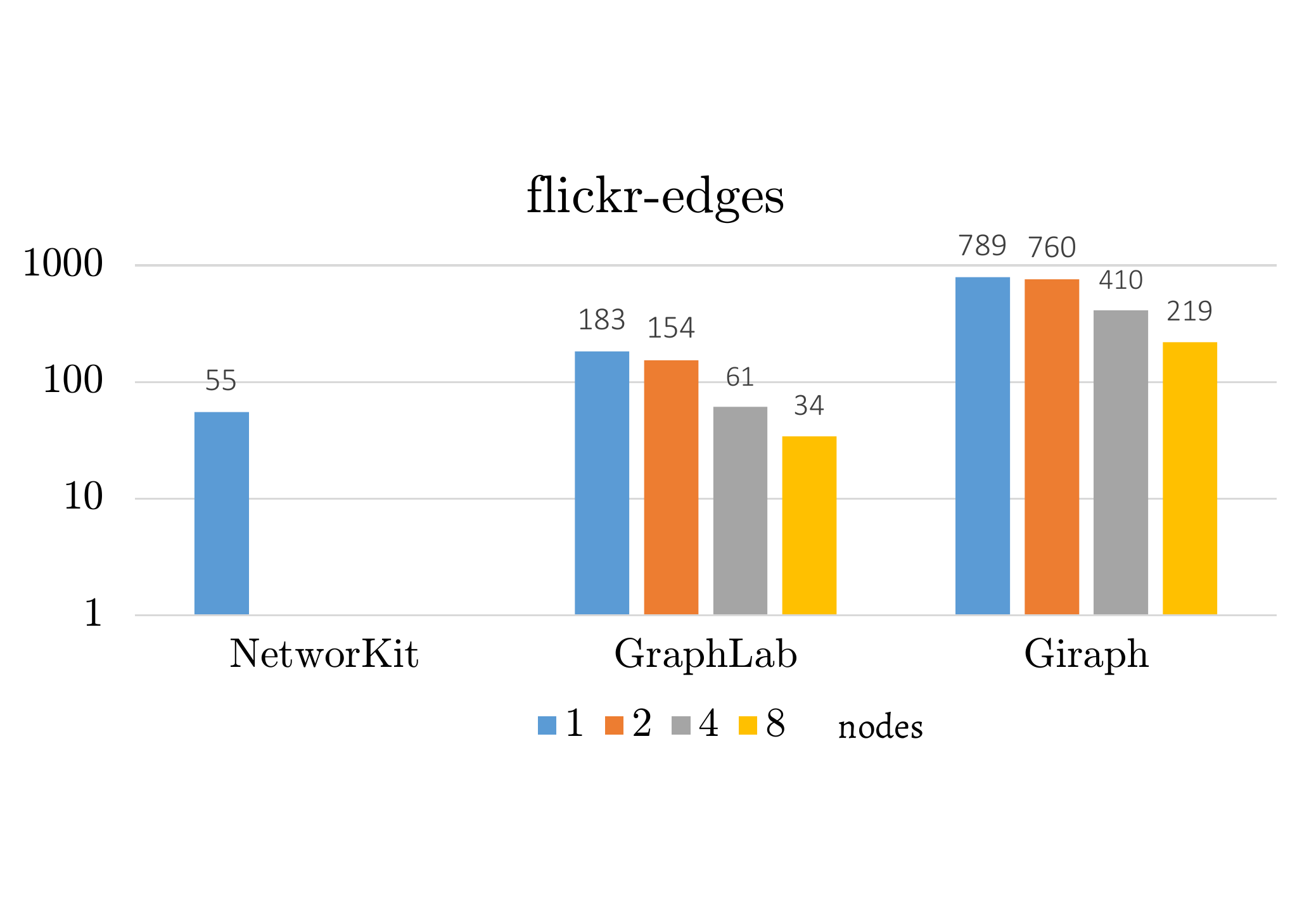}
      \caption[Clustering Coefficients running times]{Running times [in seconds] of the \emph{exact} clustering coefficient for the \graFlickrEdges graph.}
    	\label{fig:clustering_exact}
  \end{figure}

\begin{figure}[h]
  \centering
  \includegraphics[trim=0 215 108 0,clip, width=.5\textwidth]{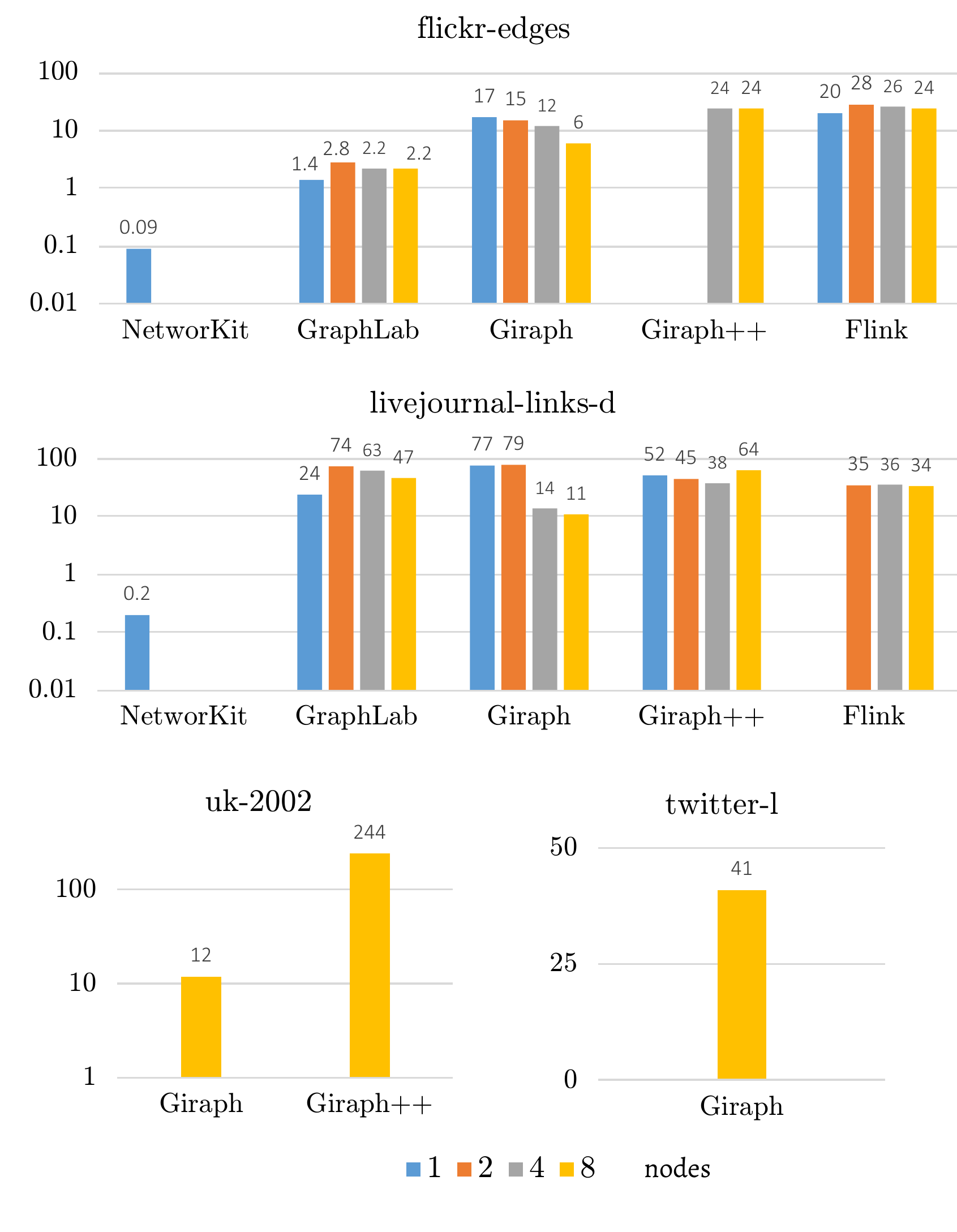}
    \caption[Clustering Coefficients approximation running times]{Running times [in seconds] for the \emph{approximate} average local clustering coefficient using 100.000 samples.}
    \label{fig:perf_clustcoeff}
  \centering
\end{figure}

The exact local clustering coefficients algorithm is the most challenging algorithm for the frameworks. It needs a lot of memory, as every vertex exchanges its total neighborhood with every neighbor, so the number of messages sent is in $O(n \cdot d_{max} ^2)$. Of the test data sets, only the small \graFlickrEdges graph can be handled by NetworKit, GraphLab, and Apache Giraph, even when the distributed frameworks use all 8 servers (see Figure~\ref{fig:clustering_exact}). Apache Flink runs the algorithm for two hours, after which the execution is aborted.
NetworKit has again the fastest execution time on a single node, followed by GraphLab and Apache Giraph. 
GraphLab needs 8 nodes to be faster than NetworKit, and Apache Giraph still takes four times as long on 8 nodes.

As expected, the approximation of the clustering coefficients allows much larger graphs to be analyzed (see Figure~\ref{fig:perf_clustcoeff}).
NetworKit's implementation is by far the fastest as it can access the graph structure much faster, while in distributed frameworks this requires message exchange, possibly over the network.
In most cases, adding additional cluster nodes does not benefit the running time (with the exception of the Giraph implementation).

\paragraph*{Memory Usage.\,}

The memory footprint of the frameworks is crucial to their use in practice -- all the more as we often experience a rather poor
scaling behavior in terms of running time for the graph computations we investigate here. Factors important for memory usage
are the underlying programming language, data structures, and the framework structure and functionality.
In Apache Giraph for example, messages sent in a vertex program are cached and only sent at the end of a superstep, which requires additional memory.

GraphLab can only process graphs which can be loaded into memory entirely, i.e. up to  \graTwitter. 
Giraph, on the other hand, offers an optional graph swapping functionality.
With this feature enabled, Giraph is able to load the \graTwitterL graph, but only the memory efficient algorithms PageRank and ConnectedComponents can be executed without memory errors.
Swapping functions naturally slow down algorithm execution due to the I/O operations.
Out of the test data sets, the largest graph Giraph++ can load on 8 nodes is \graUk. 
Apache Flink's engine also swaps out data to disk when memory runs out. In this experimental setting, it can process graphs up to the size of \graTwitter.
NetworKit is able to load the test graphs up to the size of \graUk on a single machine.

\begin{figure}[h]
  \centering
  \includegraphics[trim=0 110 90 60,clip, width=1\columnwidth]{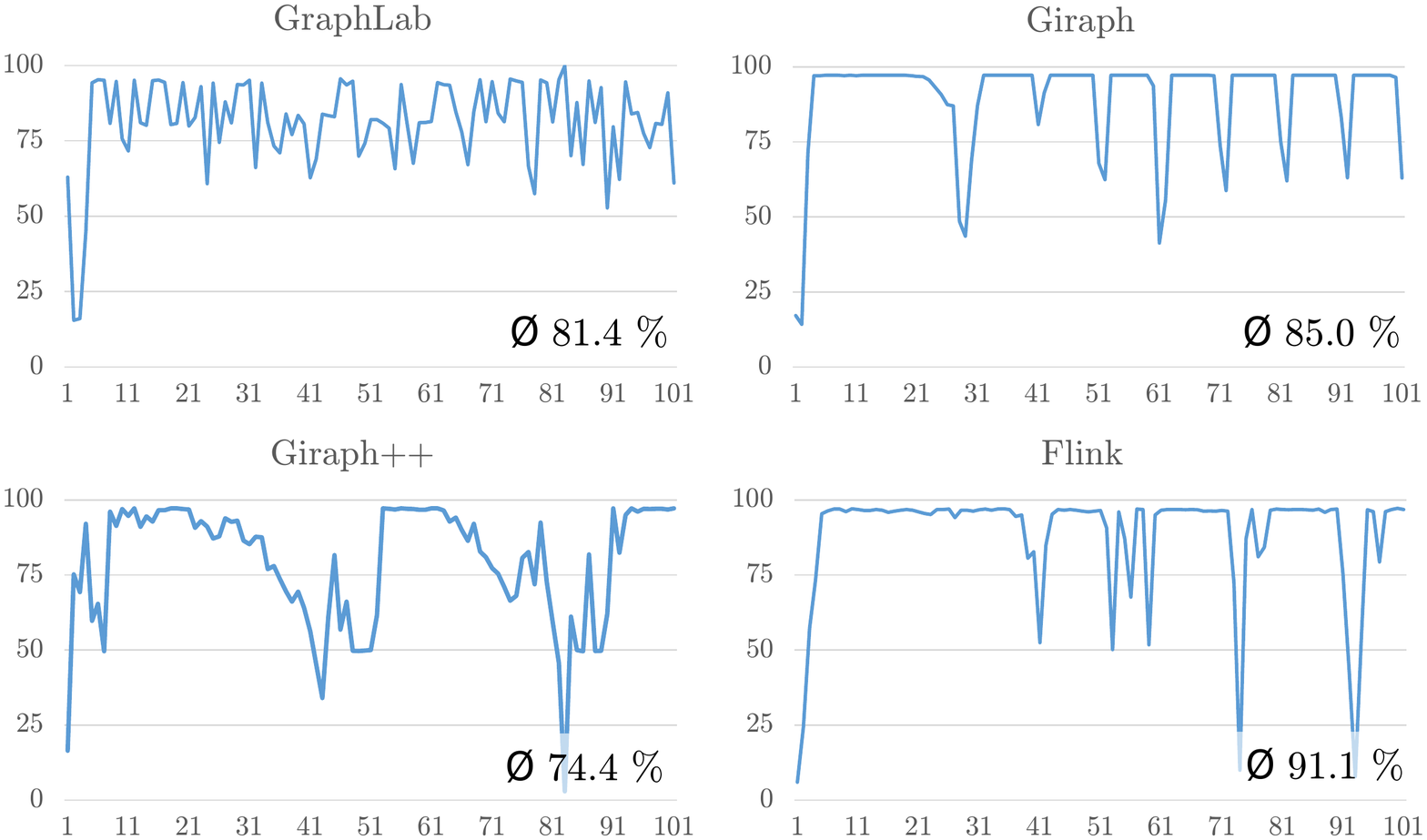}
  \caption[CPU utilization of PageRank]{CPU utilization of the first 100 seconds of the PageRank execution on the \graUk graph. The average values are taken from the whole computation.}
  \label{fig:cpu_pagerank}
  \centering
  \includegraphics[trim=0 101 90 30,clip, width=1\columnwidth]{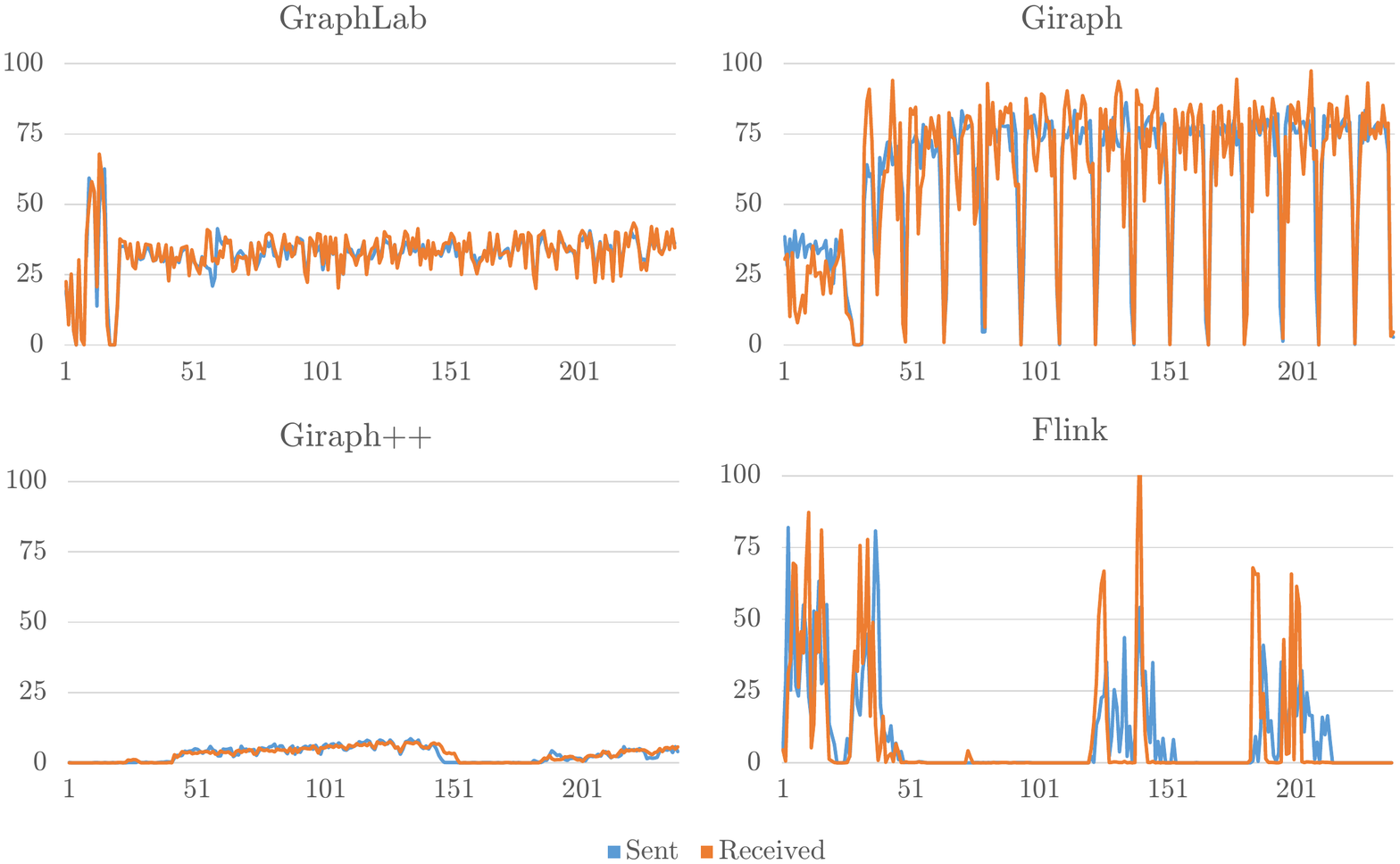}
  \caption[Network utilization of PageRank]{Network utilization of the first four minutes of the PageRank execution on the \graUk graph. The values give the send and received number of kB/s in thousands.}
  \label{fig:network_pagerank}
\end{figure}

\paragraph*{CPU and Network Utilization.\,}
In the following, the CPU and network utilization of the distributed frameworks is examined on the basis of the PageRank algorithm executed on the \graUk graph. The values were obtained using Sysstat utilities, which monitors CPU values with a time resolution of one second.
Figure \ref{fig:cpu_pagerank} shows the CPU utilization during the execution of the algorithm, starting after the I/O phase.  Iterations are visible most clearly in the Giraph execution, with low CPU activity during synchronization phases between supersteps. On average, Flink has the highest CPU utilization with over 90\%, followed by Giraph with 85\% and GraphLab with 81.4 \%. Giraph++ comes last with 74.4 \%.

The network utilization is shown in Figure \ref{fig:network_pagerank}, which gives the send and receive rates of network communication in kB/s.
When comparing the vertex-centric frameworks GraphLab and Giraph, it is noticeable that GraphLab's network communication is more evenly distributed. This can be explained with the GAS model with its decomposition in three phases, each of which requires the exchange of data for synchronization. Furthermore, GraphLab sends network messages as they arrive and not at the end of each superstep. Apache Giraph, in contrast, buffers messages during the vertex program, and sends them collectively at the end of each superstep. Therefore, there is no communication during the execution of a vertex program.
Giraph++ stands out with extraordinarily low values. The peak rate of both curves is only at around 8500 kB/s. This explains the long synchronization times and the weak performance in general of Giraph++'s executions. In the diagram of Apache Flink, the longer execution time of iterations are visible.

\paragraph*{Tuning GraphLab.\,}

Since GraphLab has emerged as the generally fastest distributed framework, we examine some of its features and tuning
options in more detail.

GraphLab's asynchronous execution engine provides opportunities for a faster algorithm convergence, but also requires synchronization to ensure serializability. The execution of \algpr and \algcc shows that the asynchronous engine indeed leads to faster convergence: The top graphs in Figure~\ref{fig:asnyc_vs_sync} show the total number of updated vertices during algorithm execution for the \graLiveU graph. The number of vertex updates in the asynchronous execution
is consistently lower than in the synchronous execution (only about 30\% to 65\%).

\begin{figure*}[h]
  \centering
\subfloat[][]{
  \begin{tikzpicture}
  \begin{axis}[
  	  title={\algcc},
      ybar,	
      bar width=8pt,
      enlargelimits=0.20,
      legend style={at={(0.5,-0.15)},
        anchor=north,legend columns=-1},
      ylabel={vertex updates in millions},
      symbolic x coords={1,2,4,8},
      xtick=data,
      xtick pos=left,
      x=1.05cm,
      area legend,
      ]
  \addplot coordinates {(1,23.926633) (2,23.926633) 	(4,23.926633) (8,23.926633)};
  \addplot coordinates {(1,7.183493) (2,	12.030581) 	(4,12.358260) (8,12.661198)};
  \legend{synch, asynch}
  \end{axis}
  \end{tikzpicture}
}
\quad
\subfloat[][]{
    \begin{tikzpicture}
    \begin{axis}[
      	title={\algccm},
        ybar,
        bar width=8pt,
        enlargelimits=0.20,
        legend style={at={(0.5,-0.15)},
          anchor=north,legend columns=-1},
        ylabel={},
        symbolic x coords={1,2,4,8},
        xtick=data,
        x=1.05cm,
        area legend,
        ]
    \addplot coordinates {(1,29.125113) (2,29.125113) 	(4,29.125113) (8,29.125113)};
    \addplot coordinates {(1,12.661962) (2,	18.300533) 	(4,19.186572) (8,17.229823)};
    \legend{synch, asynch}
    \end{axis}
    \end{tikzpicture}
}
\quad
\subfloat[][]{
    \begin{tikzpicture}
    \begin{axis}[
        ybar,
        title={\algpr},
        bar width=8pt,
        enlargelimits=0.20,
        legend style={at={(0.5,-0.15)},
          anchor=north,legend columns=-1},
        ylabel={},
        symbolic x coords={1,2,4,8},
        xtick=data,
        x=1.05cm,
        area legend,
        ]
    \addplot coordinates {(1,10.699778) (2,10.699778) 	(4,10.699778) (8,10.699778)};
    \addplot coordinates {(1,5.577698) (2,5.636313) 	(4,5.653858) (8,5.631472)};
    \legend{synch, asynch}
    \end{axis}
    \end{tikzpicture}
}
\newline
\subfloat[][]{
  \begin{tikzpicture}
  \begin{axis}[
      ybar,
      bar width=8pt,
      enlargelimits=0.20,
      legend style={at={(0.5,-0.15)},
        anchor=north,legend columns=-1},
      ylabel={Time in s},
      symbolic x coords={1,2,4,8},
      xtick=data,
      x=1.05cm,
      nodes near coords,
      nodes near coords align={vertical},
      area legend,
      ]
  \addplot coordinates {(1,27) (2,26) 	(4,23) (8,21.3)};
  \addplot coordinates {(1,24) (2,	45) 	(4,34) (8,26.9)};
  \legend{synch, asynch}
  \end{axis}
  \end{tikzpicture}
 }
\quad
\subfloat[][]{
    \begin{tikzpicture}
    \begin{axis}[
        ybar,
        bar width=8pt,
        enlargelimits=0.20,
        legend style={at={(0.5,-0.15)},
          anchor=north,legend columns=-1},
        ylabel={},
        symbolic x coords={1,2,4,8},
        xtick=data,
        x=1.05cm,
        nodes near coords,
        nodes near coords align={vertical},
        area legend,
        ]
    \addplot coordinates {(1,25) (2,24) 	(4,21) (8,17.6)};
    \addplot coordinates {(1,44) (2,	57) 	(4,45) (8,33.0)};
    \legend{synch, asynch}
    \end{axis}
    \end{tikzpicture}
 }
\quad
\subfloat[][]{
    \begin{tikzpicture}
    \begin{axis}[
        ybar,
        bar width=8pt,
        enlargelimits=0.20,
        legend style={at={(0.5,-0.15)},
          anchor=north,legend columns=-1},
        ylabel={},
        symbolic x coords={1,2,4,8},
        xtick=data,
        x=1.05cm,
        nodes near coords,
        nodes near coords align={vertical},
        area legend,
        ]
    \addplot coordinates {(1,37) (2,27) 	(4,9) (8,9.3)};
    \addplot coordinates {(1,40) (2,33) 	(4,11) (8,12.3)};
    \legend{synch, asynch}
    \end{axis}
    \end{tikzpicture}
}
    \caption[Execution time and vertex updates in asynchronous and synchronous execution]{Comparison of Graphlab's synchronous and asynchronous execution for Connected Components, message based Connected Components and PageRank (from left to right) on the \graLive graph. The top diagrams shows the number of updated vertices, the middle and bottom ones the algorithm exection time.}
    \label{fig:asnyc_vs_sync}
\end{figure*}

However, the increased convergence does not translate into faster execution times: The synchronous execution is almost always faster than the asynchronous execution. The difference is moderate with \algpr, but drastic with \algccm, where the execution time is 1.7 to 2.4 times slower than the synchronous execution time. 
What is also noteworthy is that some algorithms are faster when executed on a single machine than with two or four nodes. If executed on more than one node, GraphLab needs to maintain copies of vertices on several machines and also coordinate the execution of a vertex program to ensure serializability. This overhead can exceed the advantage of the increased computational power, thus leading to higher execution times. This behavior depends on both the algorithm and the input graph.

GraphLab provides a message API that allows vertices to send messages to other vertices, just like in the Pregel model. Furthermore, GraphLab offers delta caching which caches a vertex's gather result and updates it using delta values from neighbors. The impact of those two GraphLab functions is examined using the Connected Components algorithm, which is run in three different versions: The standard GAS implementation, a GAS implementation with delta caching, and a message-based implementation.

Figure~\ref{fig:graphlab_options} shows the execution times of the various implementations. In the synchronous execution (top graphs), the performance differences of the implementations is relatively small (5\% - 15\%). The standard implementation is slower than the delta caching and message-based variants in almost all cases. The message based implementation is slower than the delta caching implementation on the \graLiveU graph in the single- and two-node setup, but slightly slower on four and eight nodes and on all \graOrkut graph executions.
In the asynchronous execution mode, the differences are clearer. The message-based implementation is slower than the standard GAS one in almost all cases, by up to 36.7\%.

\begin{figure*}[htb]
  \centering
\subfloat[][]{
 \begin{tikzpicture}
   \begin{axis}[
   		title={\graLiveU, sync},
       ybar,
       bar width=9pt,
       enlarge y limits={upper,value=0.2},
      legend style={at={(1.03,-1.53)},
        anchor=north,legend columns=-1},
       legend style={overlay},
       ylabel={Time in s},
       symbolic x coords={1,2,4,8},
       xtick=data,
       x=1.2cm,
       nodes near coords,
       nodes near coords align={vertical},
       area legend,
       enlarge x limits=0.2,
       x post scale=1.5,
       every node near coord/.append style={font=\small,rotate=90,anchor=west}
       ]
       
        \addplot coordinates {(1,61.7) (2,41) (4, 29.2) (8,1)};
        \addplot coordinates {(1, 55.6) (2,37.5) (4, 25.5) (8,17.6)};
        \addplot coordinates {(1,52.2) (2,37) (4, 	26.9) (8,21.03)};
           \legend{Standard, Message API, Delta Caching}
        
   \end{axis}
   \end{tikzpicture}
   }
%
%
\subfloat[][]{
   \begin{tikzpicture}
     \begin{axis}[
        	title={\graOrkut, sync},
            ybar,
            bar width=9pt,
            enlarge y limits={upper,value=0.2},
            legend style={at={(0.5,-0.15)},
              anchor=north,legend columns=-1},
            symbolic x coords={1,2,4,8},
            xtick=data,
            x=1.2cm,
            nodes near coords,
            nodes near coords align={vertical},
            area legend,
            enlarge x limits=0.2,
            x post scale=1.5,
            every node near coord/.append style={font=\small,rotate=90,anchor=west}
         ]
               
         \addplot coordinates {(1,27.4) (2,26.0) (4, 23.2) (8,1)};
	         \addplot coordinates {(1, 24.6) (2,24) (4, 20.9) (8,19.33)};
         \addplot coordinates {(1,24.6) (2,24.5) (4, 	23.9) (8,23.7)};
     \end{axis}
     \end{tikzpicture}
   }
\newline 
\subfloat[][]{
 \begin{tikzpicture}
   \begin{axis}[
       ybar,
        title={\graLiveU, async},
       bar width=9pt,
       enlarge y limits={upper,value=0.2},
       legend style={at={(1,-0.17)},
         anchor=north,legend columns=-1},
                       legend style={overlay},
       ylabel={Time in s},
       symbolic x coords={1,2,4,8},
       xtick=data,
       x=1.2cm,
       nodes near coords,
       nodes near coords align={vertical},
       area legend,
       enlarge x limits=0.2,
       x post scale=1.5,
       every node near coord/.append style={font=\small,rotate=90,anchor=west}
       ]
       
        \addplot coordinates {(1,50.2) (2,60.2) (4, 40.2) (8,26.9)};
        \addplot coordinates {(1, 79.3) (2,59.4) (4,47.2) (8,33)};
   \end{axis}
   \end{tikzpicture}
}
%
%
\subfloat[][]{
   \begin{tikzpicture}
     \begin{axis}[
            ybar,
            title={\graOrkut, async},
            bar width=9pt,
            enlarge y limits={upper,value=0.2},
            legend style={at={(0.5,-0.15)},
              anchor=north,legend columns=-1},
            symbolic x coords={1,2,4,8},
            xtick=data,
            x=1.2cm,
            nodes near coords,
            nodes near coords align={vertical},
            area legend,
            enlarge x limits=0.2,
            x post scale=1.5,
            every node near coord/.append style={font=\small,rotate=90,anchor=west}
         ]
                 \addplot coordinates {(1,24.4) (2,45) (4, 34.3) (8,27.7)};
                 \addplot coordinates {(1, 44.3) (2,56.8) (4,44.6) (8,43.4)};
     \end{axis}
     \end{tikzpicture}
   }
    	\vspace{.8cm}
	\caption[GraphLab's messaging and delta caching performance]{GraphLab's execution times of the standard, message based and delta caching versions of \algcc for the \graLiveU graph (left) and the \graOrkut graph (right) in synchronous and asynchronous modes}
	\label{fig:graphlab_options}    
\end{figure*}

\section{Conclusion}

We examined several distributed computing frameworks with a focus on graph algorithms for network analysis.
The prominent MapReduce model was discarded due to its restrictions in the context of graph algorithms.
The PACT model extends the approach of MapReduce and is more powerful, although graph algorithms are generally more difficult to express than in the graph-specific models.
On the other hand, the vertex-centric Pregel and GAS are very intuitive models for this purpose.
The graph-centric extension allows to add optimizations to vertex-centric algorithms by giving the programmer access to partition information.

We implemented representative network analysis algorithms in distributed frameworks, each of which employs one of the presented programming models.
The implementations show how the programming models can be used in practice to express typical graph algorithms for network analysis, and how existing sequential algorithms such as the Clustering Coefficient approximation can be adapted to distributed programming models.
Generally, the vertex-centric Apache Giraph and GraphLab programs are very similar and the easiest to implement.
The resulting programs in Giraph++ are slightly more verbose and complex, but can use optimizations impossible in the vertex-centric model.

Experiments were conducted on a cluster of eight nodes.
We show that this distributed setting is able to process graphs of almost two billion edges, using standard commodity hardware. Not all frameworks perform equally well: Generally, GraphLab shows the best performance.
 Apache Giraph is less efficient in most cases, but offers an out-of-core mechanism which enables it to handle larger data sets. It is the only framework that could process the \texttt{twitter-l} graph.
Giraph++ unexpectedly performs worse than the vertex-centric models.
However, it must be remembered that only an experimental version of Giraph++, based on an outdated version of Apache Giraph, is available.
Apache Flink performed worse than the vertex-centric frameworks. Also, larger graphs could not be processed for some of the algorithms due to issues with memory.
Support for asynchronous execution is essential for the label propagation community detection algorithm, since the heuristic may not converge for synchronous execution due to label oscillation.

The single-machine software package NetworKit outperforms the distributed frameworks in almost all scenarios.
It can handle larger graphs than the distributed frameworks on a single machine, and its execution times are often only a fraction of the other frameworks', due to efficient native code and lack of overhead associated with the software machinery that enables and supports distributed computing.
Consequently, the decision for a distributed computing solution for complex network analysis should come out of the necessity of massive graph data volumes that exhaust typical main memory capacity.

\subsubsection*{{\scriptsize{Acknowledgements.}}}

\begin{scriptsize}
This work is partially supported by the German Research Foundation (DFG) under grant ME~3619/3-1
within the Priority Programme 1736 \emph{Algorithms for Big Data}.
Parts of this paper have been published in preliminary form
as~\cite{DBLP:conf/asunam/KochSVM15}.
\end{scriptsize}

\newpage

\bibliographystyle{apalike}
\bibliography{Literature}

\end{document}

%% file: codes/pseudocodes_setup.tex
\usepackage[ruled,vlined,linesnumbered,norelsize]{algorithm2e}
\DontPrintSemicolon

\SetAlFnt{\small\sf}
\SetKwSty{texttt}
\SetCommentSty{emph}
\let\chapter=\section 
	
\SetKwData{Vertex}{vertex}
\SetKwData{vertex}{vertex}
\SetKwData{Edge}{edge}
\SetKwData{edge}{edge}
\SetKwData{messages}{messages}
\SetKwData{Message}{message}
\SetKwData{msg}{msg}
\SetKwData{Msg}{msg}
\SetKwData{Rank}{rank}
\SetKwFunction{sendmsg}{sendMessage}
\SetKwFunction{map}{Map}
\SetKwFunction{reduce}{Reduce}
\SetKwFunction{pact}{PACT}
\SetKwFunction{emit}{Emit}
\SetKwInOut{Input}{Input}
\SetKwInOut{Output}{Output}
\SetKwFunction{gather}{gather}
\SetKwFunction{gathersum}{gather\_sum}
\SetKwFunction{messagesum}{message\_sum}
\SetKwFunction{scatter}{scatter}
\SetKwFunction{apply}{apply}
\SetKwFunction{gatherA}{gather1}
\SetKwFunction{gathersumA}{gather\_sum1}
\SetKwFunction{messagesumA}{message\_sum1}
\SetKwFunction{scatterA}{scatter1}
\SetKwFunction{applyA}{apply1}
\SetKwFunction{gatherB}{gather2}
\SetKwFunction{gathersumB}{gather\_sum2}
\SetKwFunction{messagesumB}{message\_sum2}
\SetKwFunction{scatterB}{scatter2}
\SetKwFunction{applyB}{apply2}
\SetKwFunction{init}{init}
\SetKwFunction{signal}{signal}
\SetKwFunction{append}{append}
\SetKwFunction{main}{main}
\SetKwFunction{compute}{compute}
\SetKwFunction{combine}{combine}
\SetKwFunction{mastercompute}{masterCompute}
\SetKwFunction{sendMessageToAllNeighbours}{sendMessageToAllNghbrs}

%% file: img/plots/plots_setup.tex
\definecolor{color1}{RGB}{68, 114, 196}
\definecolor{color2}{RGB}{196, 149, 0}
\definecolor{color4}{RGB}{105, 105, 105}
\definecolor{color8}{RGB}{188, 86, 15}

\definecolor{color1f}{RGB}{68, 114, 196}
\definecolor{color2f}{RGB}{255, 192, 0}
\definecolor{color4f}{RGB}{165, 165, 165}
\definecolor{color8f}{RGB}{237, 125, 49}

\definecolor{colorNk}{RGB}{68, 114, 196}
\definecolor{colorGl}{RGB}{196, 149, 0}
\definecolor{colorGi}{RGB}{105, 105, 105}
\definecolor{colorGp}{RGB}{188, 86, 15}
\definecolor{colorFl}{RGB}{55, 118, 32}

\definecolor{colorNkf}{RGB}{68, 114, 196}
\definecolor{colorGlf}{RGB}{255, 192, 0}
\definecolor{colorGif}{RGB}{165, 165, 165}
\definecolor{colorGpf}{RGB}{237, 125, 49}
\definecolor{colorFlf}{RGB}{119, 209, 87}

\pgfplotscreateplotcyclelist{FwColors}{
{colorNk,fill=colorNkf},{colorGl,fill=colorGlf},{colorGi,fill=colorGif},{colorGp,fill=colorGpf},{colorFl,fill=colorFlf},
}
\pgfplotscreateplotcyclelist{DistFwColors}{
{colorGl,fill=colorGlf},{colorGi,fill=colorGif},{colorGp,fill=colorGpf},{colorFl,fill=colorFlf},
}
\pgfplotscreateplotcyclelist{DistFwColorsCurve}{
{colorGl},{colorGi},{colorGp},{colorFl},
}
\pgfplotscreateplotcyclelist{GiGpColors}{
{colorGi,fill=colorGif},{colorGp,fill=colorGpf},
}
\pgfplotscreateplotcyclelist{nodescolors}{
{color1,fill=color1f},{color2,fill=color2f},{color4,fill=color4f},{color8,fill=color8f},
}
\pgfplotscreateplotcyclelist{nodescolorsNoTwo}{
{color1,fill=color1f},{color4,fill=color4f},{color8,fill=color8f},
}
\pgfplotscreateplotcyclelist{nodescolorsNoOneTwo}{
{color4,fill=color4f},{color8,fill=color8f},
}
\pgfplotscreateplotcyclelist{nodescolorsNoOneTwoFour}{
{color4,fill=color4f},{color8,fill=color8f},
}

\pgfplotsset{weakchart/.style={ 
  	height=5.cm,
      bar width=10pt,
      enlargelimits=0.05,
      legend style={at={(0.5,-0.15)},
        anchor=north,legend columns=-1},
      ylabel={Time in s},
      ymin = 0,
      symbolic x coords={1,2,4,8},
      xtick=data,
      xtick pos=left,
      x=3cm,
      every axis plot/.append style={line width=1.5pt,},
      cycle list name=DistFwColorsCurve,
    } 
}

%% file: codes/ch02_pregel_cc.tex
\begin{algorithm}[h]
\footnotesize
\compute{vertex, messages}
\Begin{
	\eIf{getSuperstep() = 0}{
		vertex.component $\leftarrow$ vertex.id\;
		changed $\leftarrow$ true\;
	}{
		minMsg $\leftarrow$ getMinimum(messages)\;
		changed $\leftarrow$ false\;
		\If{minMsg < vertex.component}{
			\tcp{found smaller component}
			changed $\leftarrow$ true\;
			vertex.component $\leftarrow$ minMsg\;
		}
	}
	\If{changed}{
		\tcp{component has changed - send new id to neighbors!}
		sendToAll(vertex.component)\;
	}
}
\caption[Connected Components in Pregel]{Connected Components in Pregel model}
\label{alg:pregel_cc}
\end{algorithm}

%% file: codes/ch02_gas_cc.tex
\begin{algorithm}[h]
\footnotesize
\init{vertex}
\Begin{
	vertex.component $\leftarrow$ vertex.id\;
}
\gather{vertex, edge}
\Begin{
	\KwRet{vertex.component}
}
\gathersum{v1, v2}
\Begin{
	\KwRet{minimum(v1, v2)}
}
\apply{vertex, minimum}
\Begin{
	changed $\leftarrow$ false\;
	\If{vertex.label > minimum}{
		changed $\leftarrow$ true\;
		vertex.component $\leftarrow$ minimum\;
	}
}
\scatter{vertex, edge}
\Begin{
	\If{changed}{
		\tcp{signal to inform neighbour of change}
		signal(edge.target())\;
	}
}
\caption[Connected Components in GAS]{Connected Components in GAS model}
\label{alg:gas_cc}
\end{algorithm}

%% file: codes/ch04_pagerank_gas.tex
\begin{algorithm}[h]
\footnotesize
\tcp{gather on incoming edges}
\gather{vertex, edge}
\Begin{
	\KwRet{edge.source.score / edge.source.numOutEdges}
}
\apply{vertex, sum}
\Begin{
	newScore = $\alpha$ + (1 - $\alpha$) * sum\;
	delta $\leftarrow$ newScore - vertex.score\;
	vertex.score $\leftarrow$ newScore;
	
	\eIf{|delta| > TOLERANCE}{
		converged $\leftarrow$ FALSE\;
	}{
		converged $\leftarrow$ TRUE\;
	}
}
\tcp{scatter on outgoing edges}
\scatter{vertex, edge}
\Begin{
	\If{not converged}{
		signal(edge.target)
	}
}
\tcp{alternative scatter using GraphLab's delta caching}
\scatter{vertex, edge}
\Begin{
	\If{not converged}{
		postDelta(edge.target, delta); \tcc*[r]{post delta only if vertex has not converged}
	}
}
\caption{PageRank in GAS model}
\label{alg:pagerank_gas}
\end{algorithm}

%% file: codes/ch04_pagerank_pregel.tex
\begin{algorithm}[h!]
\SetKwData{Newscore}{newScore}
\compute{vertex, messages}
\Begin{
	\eIf{getSuperstep() = 0}{
		\vertex.\Rank $\leftarrow$ 1\;
	}{
		newScore $\leftarrow$ $\alpha$ + (1 - $\alpha$) * sum(messages)\;
		delta $\leftarrow$ newScore - vertex.rank\;
		vertex.rank $\leftarrow$ newScore\;
		
		\eIf{|delta| > TOLERANCE}{
			converged $\leftarrow$ FALSE\;
		}{
			converged $\leftarrow$ TRUE\;
		}
		setAggregatorValue(CONVERGENCE, converged)\;
	}
	\eIf{getAggregatorValue(CONVERGENCE)}{
		\tcp{algorithm converged}
		voteToHalt()
	}{
		sendMessageToAllEdges(\vertex.rank / \vertex.numEdges)\;
	}
}
\combine{msg1, msg2}
\Begin{
	\KwRet{msg1 + msg2}
}
\caption{PageRank in Pregel}
\label{alg:pagerank_pregel}
\end{algorithm}

%% file: codes/ch04_clustering_pregel.tex
\begin{algorithm}[h]
\footnotesize
\compute{vertex, messages}
\Begin{
	\If{getSuperstep() = 0}{
		\For{\edge $\in$ \vertex.edges }{
			\sendMessageToAllNeighbours{\edge.otherVertex.id}
		}
	}
	\ElseIf{getSuperstep() = 1}{
		neighborEdges $\leftarrow$ 0\;
		\For{$msg\in messages$}{
			\If{\msg $\in$ \vertex.edges}{
				neighborEdges $\leftarrow$ neighborEdges + 1\;
			}
		}
		neighborEdges $\leftarrow$ neighborEdges / 2 \tcp*[r]{each link has a message from both vertices}
		possibleEdges $\leftarrow$ vertex.numEdges * (vertex.numEdges - 1) / 2\;
		vertex.localClustering $\leftarrow$ neighborEdges / possibleEdges\;
		
		\tcp{Use aggregators for global values}
		setAggregatorValue(POSSIBLE, possibleEdges)\;
		setAggregatorValue(ACTUAL, neighborEdges)\;
		setAggreagtorValue(LOCALCC, vertex.localClustering)\;
	}
}
\mastercompute{}
\Begin{
	avgLocalCC $\leftarrow$ getAggregatorValue(LOCALCC) / numVertices\;
	globalCC $\leftarrow$ getAggregatorValue(ACTUAL) / getAggregatorValue(POSSIBLE)\;
}
\caption{Clustering Coefficients in Pregel model}
\label{alg:clustering_pregel}
\end{algorithm}

%% file: codes/ch04_clustering_gas.tex
\begin{algorithm}[h]
\footnotesize
\tcp{gather on all edges}
\gatherA{vertex, edge}
\Begin{
	\KwRet{[edge.otherVertex]}
}
\gathersumA{list1, list2}
\Begin{
	append(list1, list2)\;
}
\applyA{vertex, neighbors}
\Begin{
	vertex.neighbors = neighbors\;
}
\tcp{scatter on all edges}
\scatterA{vertex, edge}
\Begin{
	otherVertex = edge.otherVertex\;
	commonNeighborsList = intersection(vertex.neighbors, otherVertex.neighbors)\;
	edge.commonNeighbors = commonNeighborsList.size\;
}
\hrulefill \;
\gatherB{vertex, edge}
\Begin{
	\KwRet{edge.commonNeighbors}
}
\applyB{vertex, neighborEdges}
\Begin{
	vertex.neighborEdges $\leftarrow$ neighborEdges\;
	vertex.possibleEdges $\leftarrow$ vertex.numEdges * (vertex.numEdges - 1) / 2\;
	vertex.localClustering $\leftarrow$ neighborEdges / vertex.possibleEdges\;
}
\hrulefill \;
\main{graph}
\Begin{
	executeVertexPrograms(graph)\;
	avgLocalCC $\leftarrow$ mapReduce(vertex.localClustering) / numVertices\;
	globalCC $\leftarrow$ mapReduce(vertex.neighborEdges) / mapReduce(vertex.neighborEdges)\;
	
}
\caption{Clustering Coefficients in GAS model}
\label{alg:clustering_gas}
\end{algorithm}

%% file: img/plots/concomp_weak.tex
\begin{figure}[h]
	\centering
  \resizebox{.5\textwidth}{!}{
 \centering
  \begin{tikzpicture}
  \begin{axis}[weakchart]
  \addplot coordinates {(1,56) (2,112) (4,214) (8,233)};
  \addplot coordinates {(1,132) (2,141) (4,149) (8,166)};
  \addplot coordinates {(1,311) (2,389) (4,455) (8,524)};
  \addplot coordinates {(1,269) (2,320) (4,367) (8,515)};
  \legend{GraphLab, 
  Giraph, Giraph++, Flink}
  \end{axis}
  \end{tikzpicture}

  }
      \caption[Connected Components weak scaling]{Weak scaling of connected components with generated data sets of 16 million edges per node}
    \label{fig:concomp_weak}
\end{figure}